\documentclass[12pt, onecolumn,journal]{IEEEtran}

\usepackage{cite}
\usepackage{graphicx}
\usepackage{psfrag}
\usepackage{subfigure}
\usepackage{url}
\usepackage{stfloats}
\usepackage{array}

\usepackage[tbtags]{amsmath} % defines many math commands and
\usepackage{amssymb}  % get, among others, blackboard bold fonts
\usepackage{verbatim} % get comment environment, + new verbatim
\newcounter{actr}
{\begin{list}{(\alph{actr})}{\usecounter{actr}}}{\end{list}}

\newcounter{ictr}
{\begin{list}{(\roman{ictr})}{\usecounter{ictr}}}{\end{list}}

\newtheorem{thm}{Theorem}
\newtheorem{lemma}{Lemma}

\newtheorem{corol}{Corollary}

\newtheorem{defn}{Definition}
\newtheorem{fact}{Fact}
\newtheorem{remark}{Remark}
\newcommand{\defeq}{\stackrel{\Delta}{=}}

\newcommand{\mrm}{\mathrm}

\newcommand{\cC}{{\mathcal{C}}}

\newcommand{\cE}{{\mathcal{E}}}

\newcommand{\cF}{{\mathcal{F}}}

\newcommand{\bH}{{\mathbf{H}}}

\newcommand{\cN}{{\mathcal{N}}}
\newcommand{\CN}{{\mathcal{CN}}}

\newcommand{\cP}{{\mathcal{P}}}

\newcommand{\cS}{{\mathcal{S}}}

\newcommand{\cU}{{\mathcal{U}}}

\newcommand{\cW}{{\mathcal{W}}}

\newcommand{\bx}{{\mathbf{x}}}

\newcommand{\cX}{{\mathcal{X}}}

\newcommand{\cY}{{\mathcal{Y}}}
\newcommand{\by}{{\mathbf{y}}}

\newcommand{\bz}{{\mathbf{z}}}

\newcommand{\g}{\gamma}

\newcommand{\eps}{\varepsilon}

\newcommand{\cH}{{\mathcal H}}

\begin{document}
\title{Secure Broadcasting}

\author{Ashish~Khisti,~\IEEEmembership{Student~Member,~IEEE,}
        Aslan~Tchamkerten,
        and~Gregory~W.~Wornell,~\IEEEmembership{Fellow,~IEEE}% <-this % stops a space
\thanks{
        This work was supported in part by NSF under Grant No.~CCF-0515109.}% <-this % stops a space
\thanks{The authors are with the Massachusetts Institute of
Technology. Email: \{khisti,tcham,gww\}@mit.edu.  This work was
presented in part at the 44th Annual Allerton Conference on
Communication, Control and Computing, Monticello, IL, September
26-29, 2006.} }

%\markboth{Journal of \LaTeX\ Class Files,~Vol.~1,
%No.~11,~November~2002}{Shell \MakeLowercase{\textit{et al.}}: Bare
%Demo of IEEEtran.cls for Journals}

\maketitle

\begin{abstract}
Wyner's wiretap channel is extended to parallel broadcast channels
and fading channels with multiple receivers.
In the first part of the paper, we consider the setup of parallel
broadcast channels with one sender, multiple intended receivers, and
one eavesdropper. We study the situations where the sender broadcasts
either a common message or independent messages to the intended receivers. 
We derive upper and
lower bounds on the common-message-secrecy capacity, which
coincide when the users are reversely degraded. For the case of
independent messages we establish the secrecy sum-capacity when the
users are reversely degraded.

In the second part of the paper we apply our results to fading
channels: perfect channel state information of all intended
receivers is known globally, whereas the eavesdropper channel is
known only to her. For the common message case, a
somewhat surprising result is proven: a positive rate can be
achieved independently of the number of intended receivers. For
independent messages, an {\emph{opportunistic}} transmission scheme
is presented that achieves the secrecy sum-capacity in the limit of
large number of receivers. Our results are stated for a fast fading
channel model. Extensions to the block fading model are also
discussed.
\end{abstract}

%
%The lower bound is achieved by a coding scheme that, when
%specialized to the case without eavesdropper, is different than
%the scheme in \cite{elGamal:80}.


\IEEEpeerreviewmaketitle

\section{Introduction}

A number of emerging applications require a ``key distribution mechanism" to
selectively broadcast confidential messages to intended receivers. For example
in \emph{pay TV systems}, a content provider wishes to selectively broadcast a
certain program to a subset of customers who have subscribed to it. An
online key distribution mechanism would allow the service provider to
distribute a decryption key to these intended receivers while securing it from
potential eavesdroppers. The program could then be encrypted via standard
cryptographic protocols, so that only users who have access to the decryption
key could view it. Indeed, in the absence of such a mechanism, current solutions
rely on variants of traditional public key cryptography (see,
e.g.,~\cite{diffie76new}) and are vulnerable to attacks such as
piracy~\cite{FiatNaor:94}.

An information theoretic framework for perfect secrecy was
developed by Shannon \cite{shannon:49}, and the problem of
broadcasting confidential messages was originally formulated by
Wyner~\cite{wyner:75Wiretap}. Wyner considered a special broadcast channel
(also known as the wiretap channel): one sender, an intended
receiver, and one eavesdropper. He characterized the tradeoff
between the rate to the intended receiver and the equivocation at
the eavesdropper when the eavesdropper has a degraded channel
compared to the intended receiver. This formulation has been
generalized for non-degraded broadcast channels
in~\cite{csiszarKorner:78}, and applied to Gaussian channels
in~\cite{leung-Yan-CheongHellman:99}.

While the results for wire-tap channels are rather surprising in
that they show that it is possible to achieve a positive rate while
keeping the eavesdropper in near-perfect equivocation, they also
provide some disappointing facts for degraded channels, such as
Gaussian~\cite{leung-Yan-CheongHellman:99}. First, the secrecy
capacity is positive only if the eavesdropper is noisier than the
intended receiver. This may not be the case in practice. Second, in
the limit of high signal-to-noise ratio (SNR), the secrecy capacity
approaches a constant and does not exhibit a logarithmic growth with
power. Thus, physical layer secrecy comes at a price in throughput
and this may have prompted many practical cryptographic solutions to
be based upon other notions of security, such as computational
security~\cite{diffie76new}. Note, however, that such solutions
require an off-line key distribution mechanism which may not be practical
in emerging applications.

The wiretap channel has received renewed interest in some recent works that
consider a wireless environment. There the eavesdropper is
not always stronger than the intended receivers due to time variations in
channel gains. These variations in turn can be exploited to communicate
securely by transmitting to the receivers that have a strong channel. Such
coding strategies may yield a practical approach for secure communication
without an off-line key agreement.

In the present work we extend Wyner's wiretap channel to parallel
broadcast channels with one sender, multiple intended receivers,
and one eavesdropper. We consider two situations: all intended receivers
get a common message or independent messages. We first derive
upper and lower bounds on the common-message-secrecy-capacity.
These bounds coincide when the users are reversely degraded.
Perhaps the main observation is that, to achieve the common
message capacity, independent codebooks are used on each parallel
channel, and each receiver jointly decodes its received sequences. 
Next, we consider the case where the intended receivers get 
independent messages. We establish the secrecy capacity for the
reversely degraded case. The achievable scheme is simple: transmit
to the strongest user on each parallel channel and use independent
codebooks across the channels. Our results for the parallel broadcast channels
can be viewed as generalizations of the results in~\cite{elGamal:80} which considers
a similar setup without the presence of an eavesdropper.

Our study on parallel channels provides insights to the problem of
broadcasting confidential messages over fading channels. In the
second part of the paper we consider an i.i.d. fading model. We
assume the intended receivers' channel state information (CSI) is
revealed to all communicating parties (including the
eavesdropper), while the eavesdropper's channel gains are revealed
only to her.

We first examine the case when a common message needs to be
delivered to all intended receivers in the presence of potential
eavesdroppers. We refer to this problem as secure multicasting. We present a
scheme that exploits CSI at the transmitter and achieves a rate that
does not decay to zero with increasing number of receivers. Note
that, without a secrecy constraint,
transmitter CSI appears to be of little value for multicasting over
ergodic channels. Indeed the capacity appears to be not too far from
the maximum achievable rate with a flat power allocation scheme. The
secrecy constraint adds a new twist to the multicasting problem as
it requires to consider protocols that exploit transmitter CSI.

For the case of independent messages, we consider an opportunistic
scheme that selects the user with the strongest channel at each
time. We use Gaussian wiretap codebooks for each intended receiver
and show that this scheme achieves the sum capacity in the limit
of large number of receivers. Our results can be interpreted as
the wiretap analog of the multiuser diversity results in
settings without secrecy constraint (see, e.g.,~\cite{tseViswanath:05}).

In related works, the Gaussian wiretap channel was extended to
parallel channels in~\cite{yamamoto:91}. More recently, the case
of discrete memoryless parallel channels with one receiver and one
eavesdropper has been studied
in~\cite{liangPoor:06a,liYatesTrappe:06a}. The wiretap setting has
been also studied for fading channels
in~\cite{NegiGoel05,barros:06,gopalaLaiElGamal:06Secrecy}. All
these works consider the setup of one sender, one receiver, and
one eavesdropper.

There is also a vast literature on multiuser diversity in
broadcast channels with independent messages starting from the
results in~\cite{Tse:97,liGoldsmith:98}. However, to the best of
our knowledge, the present work is the first to consider
the impact of multiuser diversity on secrecy systems. As discussed
before, the case of a common message has received much less
attention in the literature. The problem of transmitting a common
message on parallel channels has been studied
in~\cite{elGamal:80,jindalGoldsmith:04} but we are not aware of a
general treatment of this problem for fading channels (without an
eavesdropper). We hope that the secrecy constraint creates renewed
interest in the study of common message broadcast problems, given
its application to key distribution.

We use the following notation. Upper case letters are used for random variables
and the lower case for their realizations. The notation $s^n$ denotes a vector
of length $n$. Vector quantities related to the eavesdropper have a subscript $e$, e.g.,
$y_e^n$, while the ones of the intended receivers are
subscripted by the user number, e.g., $y_{i}^n$. We use the subscript $i$
to index the receivers and the subscript $j$ to index  the channels. We 
use the letter $t$ to denote the discrete time index. If
there is an ordering of users on a given channel, the strongest user on channel
$j$ will be denoted by $\pi_j$. The set of ordered users on channel $j$ is
denoted as $\pi_j(1), \pi_j(2), \ldots$ We use the notation $p(X_j)$ to
denote the probability mass function of random variable $X_j$.

\section{Parallel Channels: Model}

\label{subsec:chanModel} In our setup, there are $M$ parallel
channels for communication, one sender, $K$ intended receivers, and one
eavesdropper.
\begin{defn}[Product Broadcast Channel]
An $(M,K)$ product broadcast channel consists of one sender, $K$ receivers,
one eavesdropper, and $M$ channels. The channels have finite input and output
alphabets, are memoryless and independent of each other, and are characterized
by their transition probabilities given by
\begin{equation}
\Pr\left(\{y_{1j}^n,y_{2j}^n,\ldots,
y_{Kj}^n,y_{ej}^n\}_{j=1,\ldots,M} \mid \{x_{j}^n\}_{j=1,\ldots M}
\right) = \prod_{j=1}^M \prod_{t=1}^n
\Pr(y_{1j}(t),y_{2j}(t),\ldots,y_{Kj}(t),y_{ej}(t) \mid x_{j}(t))
\label{eq:ChanIndep}
\end{equation}
for $j=1,2,\ldots,M$, where $x_j^n = x_j(1), x_j(2),\ldots, x_j(n)$
denotes the sequence of symbols transmitted on channel $j$, and where
$y_{ij}^n=y_{ij}(1),y_{ij}(2),\ldots,y_{ij}(n)$ denotes the sequence
of symbols received by user $i$ on channel $j$ from time $1$ up to
$n$. The alphabets of the $X_j$'s and $Y_{ij}$'s are denoted by
$\cX$ and $\cY$ respectively.\label{defn:ProductBroadcastChannels}
\end{defn}

Of particular interest is a special class of reversely degraded
broadcast channels.

\begin{defn}[Reversely Degraded Broadcast Channel]
An $(M,K)$ reversely degraded broadcast channel is an $(M,K)$
product broadcast channel, where each of the $M$ parallel channels
is degraded in a certain order. For some permutation
$\pi_j(1),\pi_j(2),\ldots \pi_j(K+1)$ of the set $\{1,2,\ldots, K,
e\}$ of the $K+1$-receivers, a Markov chain $X_j \rightarrow
Y_{\pi_j(1)} \rightarrow Y_{\pi_j(2)} \rightarrow \ldots
\rightarrow Y_{\pi_j(K+1)}$ can be specified. 
\label{defn:reverselyDegradedBC}
\end{defn}
\begin{figure}
\centering
\includegraphics[scale=0.5]{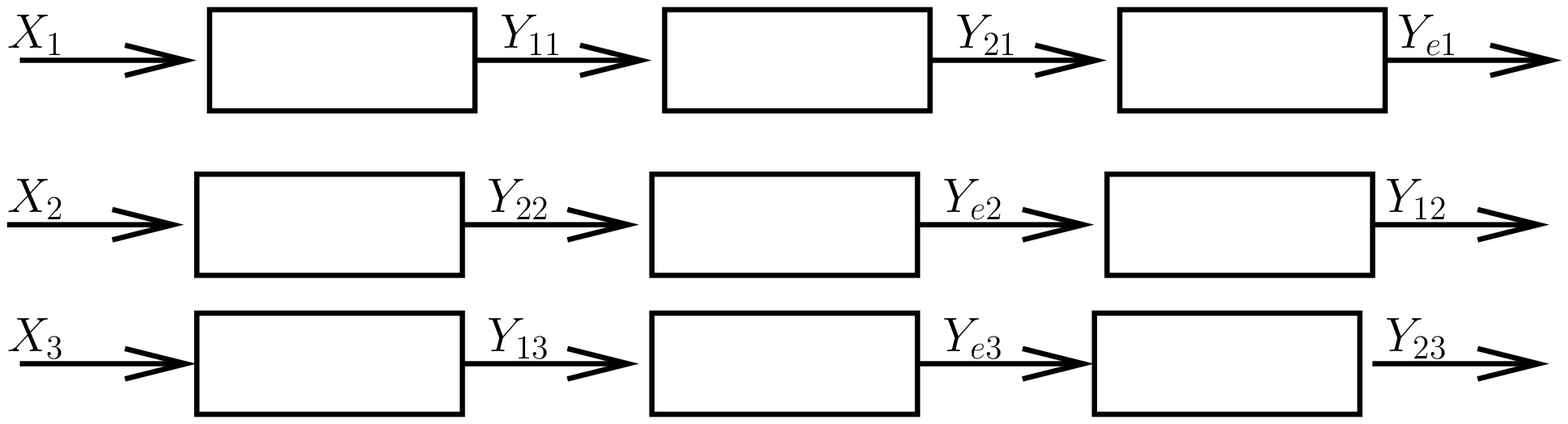}
\caption{An example of reversely degraded parallel channel in
Definition~\ref{defn:reverselyDegradedBC} with one sender, $K=2$ users, one
eavesdropper, and $M=3$ channels.} \label{fig:parchan}
\end{figure}

\begin{remark}
Note that in Definition~\ref{defn:reverselyDegradedBC} the order
of degradation can be different across the channels, so the
overall channel may not be degraded. An example of reversely
degraded parallel channel is shown in Fig~\ref{fig:parchan}. Also, on any
parallel channels component, the $K$ users and the eavesdropper  are \emph{physically} degraded. Our capacity results will,
however, only depend on the marginal distribution of receivers on
each channel (see Fact~\ref{fact:SATO} below). Accordingly, these
results also hold for a larger class of channels where receivers
on each channel are stochastically degraded.
\end{remark}

\section{Parallel Channels: Common Message}

\label{subsec:CommonMsgParallel} In this section we consider the
case where all the receivers are interested in only a common
message. This common message must be protected from the
eavesdropper in the sense described below.

\begin{defn}
A $(n,2^{nR})$ code consists of a message set $\cW = \{1,2,\ldots
2^{nR}\}$, a (possibly stochastic) mapping $\displaystyle
\omega_n:\cW \rightarrow \underbrace{\cX^n \times \cX^n \times
\ldots \times \cX^n}_{M \text{ times}} $ from the message set to the
codewords for the $M$ channels, and a decoder
$\displaystyle\Phi_{i,n}:
\underbrace{\cY^n\times\cY^n\times\ldots\ldots
\ldots\times\cY^n}_{M\text{ times}} \rightarrow \cW$ for
$i=1,2,\ldots K$ at each receiver. We denote the message estimate at decoder
$i$ by $\hat{W}_i$. A common-message-secrecy-rate $R$ is achievable
if, for any $\eps>0$, there exists a length $n$ code such that
$\Pr(W \neq \hat{W}_i) \le \eps$ for $i=1,2,\ldots K$, while
\begin{equation}
\frac{1}{n}H(W | Y_{e1}^n,Y_{e2}^n,\ldots, Y_{eK}^n) \ge R-\eps.
\label{eq:SecrecyConstCommonMsg}
\end{equation}
The common-message-secrecy-capacity is the supremum over all
achievable rates. \label{defn:CommonMessageRateDefn}
\end{defn}

\begin{remark}
\label{remark:SecrecyCapacityImp} Wyner's formulation considers the
rate-equivocation region $(R,R_e)$  with $\frac{1}{n}H(W)\ge R$ and
$\frac{1}{n}H(W|Y_e^n) \ge R_e$. The secrecy-capacity constitutes
the special case when $R=R_e$. In the key-distribution application
of interest, the key length is limited by the equivocation
rate $R_e$ --- the minimum number of bits the eavesdropper needs to
guess to decode the message. Accordingly, the secrecy capacity is of
primary interest.
\end{remark}

\subsection{Main Results}
Our main result is the
characterization of upper and lower bounds on the
common-message-secrecy-capacity for the product channel
model~\eqref{eq:ChanIndep}. The bounds coincide for the reversely
degraded model.

To state our upper bound we introduce the following additional
notation. For any $j=1,2,\ldots, M$, let $\cP_j$ denote the
collection of all joint distributions $p'(Y_{1j},Y_{2j},\ldots
Y_{Kj}, Y_{ej}| X_j)$ with the same marginal marginal distribution
as $p(Y_{1j}|X_j), p(Y_{2j}|X_j),\ldots,
p(Y_{Kj}|X_j),p(Y_{ej}|X_j)$. Let $\cP = \cP_1 \times \cP_2 \times
\ldots \times \cP_M$ denote the cartesian product of these sets across the
channels.

\begin{lemma}[Upper Bound] For the product broadcast channel model in Definition~\ref{defn:ProductBroadcastChannels},
an upper bound on the secrecy capacity is given by
\begin{equation}
 R^{+,\mrm{common}}_{K,M} \defeq
\min_\cP\max_{\prod_{j=1}^M p(X_j)} \min_{i \in \{1,2,\ldots, K\}}
\sum_{j=1}^M I(X_j;Y_{ij}|Y_{ej}) \label{eq:CommonMessageUpperBound}
\end{equation}
\label{lem:CommonMessageUpperBound} where the first minimum is
over all the joint distributions $$\{p'(Y_{1j},Y_{2j},\ldots
Y_{Kj}, Y_{ej}| X_j)\}_{j=1}^M\in \cP. $$
\end{lemma}
\begin{lemma}[Lower Bound]
An achievable common-message-secrecy-rate for the product
broadcast channel model
Definition~\ref{defn:ProductBroadcastChannels} is\footnote{$\{v\}^+$ stands for $\max\{0,v\}$.}
\begin{equation}
 R^{-,\mrm{common}}_{K,M}
\defeq \max_{\substack{\prod_{j=1}^M p(U_j)\\ \{X_j=
f_j(U_j)\}_{j=1,\ldots,M}}} \min_{ i \in \{1,2,\ldots,K\}}
\sum_{j=1}^M \{I(U_j;Y_{ij}) - I(U_j; Y_{ej})\}^+\;.
\label{eq:CommonMessageLowerBound}
\end{equation}
The random variables $U_1,U_2, \ldots U_M$ are independent over
some alphabet $\cU$, and each $f_j: \cU\rightarrow \cX$,
$j=1,\ldots,M$ is a (possibly stochastic\footnote{For each $u\in
\cU_j$, a stochastic mapping $f_j:\cU\rightarrow X$ produces a
random element in $\cX$.}) mapping from the $\cU$ to $\cX$.

\label{lem:CommonMessageLowerBound}
\end{lemma}

Our upper and lower bounds coincide for the case of reversely
degraded product channels.
\begin{thm}
\label{thm:CommonMsgCap} The common-message-secrecy-capacity for
the reversely degraded channel model in
Definition~\ref{defn:reverselyDegradedBC} is given by
\begin{equation}
C^\mrm{common}_{K,M} =\max_{\prod_{j=1}^M p(X_j)} \min_{ i \in
\{1,2,\ldots,K\}} \sum_{j=1}^M I(X_j;Y_{ij}| Y_{ej})\;.
\label{eq:CommonMsgCap}
\end{equation}
\end{thm}
Note that the expression in~\eqref{eq:CommonMsgCap} is evaluated
for the joint distribution induced by the reversely degraded
channel. This distribution is the worst-case distribution in the
set $\cP$ in~\eqref{eq:CommonMessageUpperBound}.

\begin{remark}
Our achievable rate expression in~\eqref{eq:CommonMessageLowerBound}
involves optimization over the auxiliary random variables $U_j$ and
the stochastic mappings $f_j(\cdot)$. As noted
in~\cite{csiszarKorner:78}, the expression $I(U_j;Y_{ij})-
I(U_j;Y_{ej})$ is in general not convex in $p(X_j|U_j)$, hence the
optimal $f_j(\cdot)$ need not be deterministic functions. However,
for the special reversely degraded case in Theorem~\ref{thm:CommonMsgCap},
the choice $X_j = U_j$ is optimal (see
Section~\ref{subsec:proofThm1}).
\end{remark}

The proof of the upper bound in Lemma~\ref{lem:CommonMessageUpperBound} is a rather straightforward extension of Wyner's converse for the single user wiretap channel. The achievability proof in Lemma~\ref{lem:CommonMessageLowerBound} is more interesting. When specialized to the case of no eavesdropper, it provides a different capacity achieving scheme than the one considered in~\cite{elGamal:80}.

\subsection{Upper Bound}
%In this section, we provide a detailed
%proof for the upper bound~\eqref{eq:CommonMessageUpperBound}.

\begin{fact}
\label{fact:SATO} The common-message-secrecy-capacity for the
wiretap channel depends only on the marginal distributions
$p(Y_{1j}|X_j), p(Y_{2j}|X_j),\ldots, p(Y_{Kj}|X_j)$
in~\eqref{eq:ChanIndep} and not on the joint distribution
$p(Y_{1j},Y_{2j},\ldots,Y_{Kj}|X_j)$ for each $j=1,2,\ldots, M$.
\end{fact}
The proof of this fact is essentially the same as the proof for 
broadcast channels without secrecy constraint (see, e.g.,
\cite{csiszarKorner:78}).

The following property will be used in the upper bound derivation
but, also, in other subsequent proofs.
\begin{fact}
\label{fact:concavity} For any random variables $X$, $Y$, and $Z$
the quantity $I(X;Y|Z)$ is concave in $p(X)$.
\end{fact}
The proof is implicit in the arguments in~\cite{Dijk97}. We provide
it in Appendix~\ref{app:concavityProof} for completeness.

Suppose there exists a sequence of $(n,2^{nR})$ codes such that,
for every $\eps>0$, as $n\rightarrow\infty$
\begin{equation}
\begin{aligned}
&\Pr(W \neq \hat{W}_i) \le \eps, \quad i=1,2,\ldots K\\
&\frac{1}{n}I(W;Y_{e1}^n,\ldots, Y_{eM}^n) \le \eps.
\end{aligned}
\label{eq:AchievabilityConditionCommonMessage}
\end{equation}

We first note that from Fano's Lemma we have
\begin{equation}
\frac{1}{n}H(W|Y_{i1}^n, Y_{i2}^n,\ldots, Y_{iM}^n) \le
\frac{1}{n} + \eps R\quad i=1,2,\ldots K. \label{eq:Com_Conv_Fano}
\end{equation}

Combining ~\eqref{eq:AchievabilityConditionCommonMessage}
and~\eqref{eq:Com_Conv_Fano} we have, for all $i=1,2,\ldots K$ and
$\eps' = \eps + \frac{1}{n} + \eps R$,
\begin{align}
nR &\le I(W;Y_{i1}^n,\ldots,Y_{iM}^n) -
I(W;Y_{e1}^n,\ldots,Y_{eM}^n) + n\eps' \nonumber \\
&\le I(W;Y_{i1}^n,\ldots,Y_{iM}^n |
Y_{e1}^n,\ldots,Y_{eM}^n) + n\eps'\nonumber\\
&= h(Y_{i1}^n,\ldots, Y_{iM}^n|Y_{e1}^n,\ldots,Y_{eM}^n) -
h(Y_{i1}^n,\ldots, Y_{iM}^n|Y_{e1}^n,\ldots,Y_{eM}^n,
W)\nonumber\\
&\le h(Y_{i1}^n,\ldots, Y_{iM}^n|Y_{e1}^n,\ldots,Y_{eM}^n) -
h(Y_{i1}^n,\ldots,
Y_{iM}^n|Y_{e1}^n,\ldots,Y_{eM}^n,X_1^n,\ldots,X_M^n,
W)\nonumber\\
&=h(Y_{i1}^n,\ldots, Y_{iM}^n|Y_{e1}^n,\ldots,Y_{eM}^n) -
h(Y_{i1}^n,\ldots,
Y_{iM}^n|Y_{e1}^n,\ldots,Y_{eM}^n,X_1^n,\ldots,X_M^n)\label{eq:Com_Conv_Markov}\\
% &= I(X_1^n,\ldots X_M^n ; Y_{i1}^n,\ldots,Y_{iM}^n |
%Y_{e1}^n,\ldots,Y_{eM}^n) + n\eps' \nonumber\\
&= h(Y_{i1}^n,\ldots,Y_{iM}^n | Y_{e1}^n,\ldots,Y_{eM}^n) -
\sum_{j=1}^M h(Y_{ij}^n|X_j^n, Y_{ej}^n) +n\eps'
\label{eq:Com_Conv_Indep}\\
&\le \sum_{j=1}^M h(Y_{ij}^n | Y_{ej^n})- \sum_{j=1}^M
h(Y_{ij}^n|X_j^n, Y_{ej}^n) +n\eps'
\nonumber\\
 &\le \sum_{j=1}^M I(X_j^n; Y_{ij}^n | Y_{ej}^n) +
n\eps',\label{eq:Com_last}
\end{align}
where~\eqref{eq:Com_Conv_Markov} follows from the fact that
 $W \rightarrow (X_1^n,\ldots X_M^n, Y_{e1}^n,\ldots, Y_{eM}^n) \rightarrow
 (Y_{i1}^n,\ldots,Y_{iM}^n)$ form a Markov chain, and
 ~\eqref{eq:Com_Conv_Indep} holds because the parallel
 channels are mutually independent in~\eqref{eq:ChanIndep} so that
 $$h(Y_{i1}^n,\ldots,Y_{iM}^n | Y_{e1}^n,\ldots,Y_{eM}^n, X_1^n,\ldots,X_M^n)
 =
\sum_{j=1}^M h(Y_{ij}^n|X_j^n, Y_{ej}^n)\;.
$$ We now upper bound each term in the summation \eqref{eq:Com_last}.
We have
\begin{align}
I(X_j^n; Y_{ij}^n | Y_{ej}^n) &\le \sum_{k=1}^n I(X_j(k);
Y_{ij}(k)
| Y_{ej}(k))\label{eq:Com_Conv_Memoryless}\\
&=\sum_{k=1}^n I(X_j(k); Y_{ij}(k) ,Y_{ej}(k)) -
I(X_j(k);Y_{ej}(k))\\
&=nI(X_j; Y_{ij} ,Y_{ej}|Q) -
nI(X_j;Y_{ej}|Q)\label{eq:Com_Conv_Thm2TS}\\
&= n I(X_j; Y_{ij}|Y_{ej}, Q) \nonumber\\
&\le n I(X_j; Y_{ij} | Y_{ej}) \label{eq:Com_Conv_Concave},
\end{align}
where~\eqref{eq:Com_Conv_Memoryless} follows from the fact that
the channel is memoryless,and ~\eqref{eq:Com_Conv_Thm2TS} is obtained
by defining $Q$ to be a (time-sharing) random variable uniformly
distributed over $\{1,2,\ldots, n\}$ independent of everything
else. The random variables $(X_j,Y_{ij},Y_{ej})$ are such that,
conditioned on $Q =k$, they have the same joint distribution as
$(X_j(k),Y_{ij}(k),Y_{ej}(k))$.
Finally~\eqref{eq:Com_Conv_Concave} follows from the fact that the
mutual information is concave with respect to the input
distribution $p(X_j)$ as stated in Fact~\ref{fact:concavity}.

Combining~\eqref{eq:Com_Conv_Concave}
and~\eqref{eq:Com_Conv_Indep} we have
\begin{align}
R &\le  \sum_{j=1}^M I(X_j; Y_{ij} | Y_{ej}) + \eps', \quad
i=1,2,\ldots K \nonumber\\
&= \min_{1\le i \le K} \sum_{j=1}^M I(X_j; Y_{ij} | Y_{ej}) +
\eps' \\
&\le \max_{\prod_{j=1}^M p(X_j)} \min_{1\le i \le K} \sum_{j=1}^M
I(X_j; Y_{ij} | Y_{ej}) + \eps'\label{eq:Com_Conv_UBNoSato}\;.
\end{align}

The above bound~\eqref{eq:Com_Conv_UBNoSato} depends on the joint
distribution across the channels. Accordingly, we tighten the upper
bound by considering the worst distribution in $\cP =
\cP_1\times\cP_2\times\ldots\times\cP_M$ which gives
\begin{align}
R &\le  \min_{\cP}\max_{\prod_{j=1}^M p(X_j)}\min_{1\le i \le K}
\sum_{j=1}^M I(X_j; Y_{ij} | Y_{ej}) + \eps'\;.
\label{eq:Com_Conv_UBNoSato2}
\end{align}

\subsection{Lower Bound}

\begin{figure}
\centering
\includegraphics[scale=0.7]{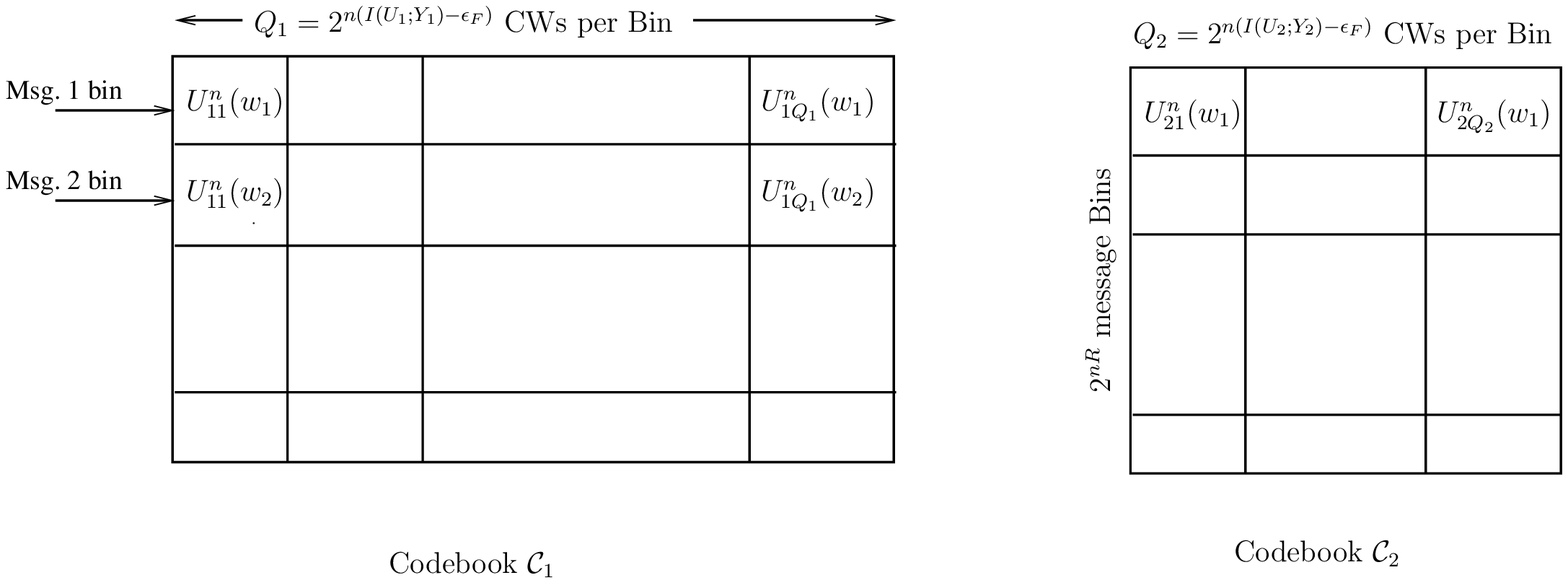}
\caption{Structure of the codebooks in our coding scheme for the
case of two parallel channels. Each codebook has $2^{nR}$ message
bins and $ Q_j \approx 2^{n(I(U_j;Y_{ej}))}$ codewords per message
bin. Thus the size of bins depends on the mutual information of
the eavesdropper on the corresponding channel. This flexible
binning enables to confuse the eavesdropper on each channel.
Note that $\cC_1$ and $\cC_2$ above have the same number of rows
but different number of columns. The codewords for message $w_k$
in $\cC_j$ are labeled as $u_{j1}^n{(w_k)}, \ldots
u_{jQ_j}^n(w_k)$}. \label{fig:CodebookDesign}
\end{figure}

We first informally present the main ideas in our achievability scheme. We
construct $M$ independent codebooks, one for each channel, denoted
as $\cC_1, \cC_2,\ldots, \cC_M$. The structure of the codebooks is
shown in Fig.~\ref{fig:CodebookDesign}. Each $\cC_j$ has $2^{n{(R+
I(U_j;Y_{ej}))}}$ codewords, randomly partitioned into
$2^{nR}$ \emph{message bins} --- there are $2^{nI(U_j;Y_{ej})}$
codewords per bin. Given a message $W$, the encoder selects $M$
codewords as follows. On channel $j$, it looks into the bin
corresponding to message $W$ in $\cC_j$ and randomly selects a
codeword in this bin. Each intended receiver attempts to find a
message that is jointly typical with its received sequences. An
appropriate choice of  $R$ 
guarantees successful decoding with high probability for each
intended receiver, and near perfect equivocation at the
eavesdropper.

We now provide a formal description of our coding scheme.

Fix the distributions $p(U_1), p(U_2),\ldots, p(U_M)$ and the
(possibly stochastic) functions $f_1(\cdot),\ldots, f_M(\cdot)$. Let
$\eps_E$ and $\eps_R$ be positive constants, to be quantified later.
With respect to these quantities, define
\begin{equation}
\begin{aligned}
R &= \min_{1\le i \le K }\sum_{j=1}^M \{I(U_j;Y_{ij}) - I(U_j;
Y_{ej})\}^+ - \eps_R\\
R_{ej} &= I(U_j;Y_{ej}) - \eps_F, \quad j=1,2,\ldots M.
\end{aligned}
\label{eq:Com_Achiev_R_Defn}
\end{equation}

In what follows, whenever typicality is mentioned it is intended to
be $\eps-$weak typicality (see, e.g.,~\cite{coverThomas}). The set
$T(U_j)$ denotes the set of all sequences that are typical with
respect to distribution $p(U_j)$ and the set $T(X_j,U_j)$ denotes
the set of all jointly typical sequences $(x_j^n,u_j^n)$ with
respect to the distribution $p(X_j,U_j)$. $T_{u_j^n}(X_j|U_j)$
denotes the set of all sequences $x_j^n$ conditionally typical with
respect to a given sequence $u_j^n$ according to $p(X_j|U_j)$.

\subsubsection{Codebook Generation}
\begin{itemize}

\item Codebook $\cC_j$ for $j=1,2,\ldots, M$ has a total of $M_{j}
= 2^{n(R + R_{ej})}$ length $n$ codeword sequences. Each sequence is
selected uniformly and independently from the set $T(U_j)$.

\item We randomly partition the $M_j$ sequences into $2^{nR}$
message bins so that there are $Q_j = 2^{nR_{ej}}$ codewords per
bin.

\item The set of codewords associated with bin $w$ in codebook
$\cC_j$ is denoted as
\begin{equation}
\cC_j(w) = \{u^n_{j1}(w),u^n_{j2}(w),\ldots, u^n_{jQ_j}(w),\},
\quad w=1,2,\ldots 2^{nR}, \quad j=1,2,\ldots M.
\end{equation}
Note that $\cC_j = \bigcup_{w=1}^{2^{nR}}\cC_j(w)$ is the codebook
on channel $j$.
\end{itemize}

\subsubsection{{Encoding}}
To encode message $w$, the encoder randomly and uniformly selects
a codeword in the set $\cC_j(w)$ for all $1\le j \le M$.
Specifically,
\begin{itemize}

\item Select $M$ integers $k_1,k_2,\ldots,k_M$, where $k_j$ is
selected independently and uniformly from the set $\{1,2,\ldots
Q_j\}$.
\item Given a message $w$, select a codeword
$u^n_{jk_j}(w)$ from codebook $\cC_j(w)$ for $j=1,2,\ldots M$.

\item The transmitted sequence on channel $j$ is denoted by
$x_j^n=x_j(1),x_j(2),\ldots,x_j(n)$.  The symbol $x_j(t)$ is
obtained by applying the (possibly stochastic) function
$f_j(\cdot)$ on the $t^\mrm{th}$ element of the codeword
$u^n_{jk_j}(w)$.
\end{itemize}

\subsubsection {Decoding}
 Receiver $i$, based on its observations  $(y_{i1}^n, y_{i2}^n, \ldots, y_{iM}^n)$
 from the $M$ parallel channels, declares message $w$ according to the following rule.
\begin{itemize}
\item Let $\cS_i =\{j | 1\le j\le M, I(U_j;Y_{ij})> I(U_j;Y_{ej})\}$ denote the set of channels where receiver $i$ has
larger mutual information than the eavesdropper. The receiver only
considers the outputs $y_{ij}^n$ from these channels.

\item Receiver $i$ searches for a message $w$ such that, for each
$j\in \cS_i$, there is an index $l_j$ such that $(u_{jl_j}^n(w),
y_{ij}^n) \in T(U_{j},Y_{ij})$. If a unique $w$ has this property,
the receiver declares it as the transmitted message. Otherwise, the
receiver declares an arbitrary message.
\end{itemize}

\subsubsection{Error Probability}
We show that, averaged over the ensemble of codebooks, the error
probability is smaller than a constant $\eps'$ (to be specified),
which approaches zero as $n\rightarrow \infty$. This demonstrates
the existence of a codebook with error probability less than
$\eps'$. We do the analysis for user $i$ and, without loss of
generality, assume that message $w_1$ is transmitted.
\begin{itemize}

\item {\bf False Reject Event}: Let $\cE_{1j}^c$ be the event
$\{(U_{jk_j}^n(w_1),Y_{ij}^n)\notin T(U_j, Y_{ij})\}$. Since $U_j^n
\in T(U_j)$ by construction and $Y_{ij}$ is obtained by passing
$U_j$ through a DMC, it follows that $\Pr(\cE_{1j}^c)\le \delta$,
where $\delta \rightarrow 0$ as $\eps \rightarrow 0$. Accordingly if
$\cE_1^c$ denotes the event that message $w_1$ does not appear
typical, then we have
\begin{equation}
\Pr(\cE_1^c) = \Pr\left(\bigcup_{j=1}^M \cE_{1j}^c \right)\le
M\delta.
\end{equation}
\item {\bf False Accept Event}: As before, let $\cS_i \subseteq
\{1,2, \ldots,M\}$ denote the subset of channels for which
$I(U_j;Y_{ij})> I(U_j;Y_{ej})$. In what follows the index $j$ will
only refer to channels in $\cS_i$.

Let $\cE_{rj}$ denote the event that there is a codeword in the
set $\cC_j(w_r)$ ($r
> 1$) typical with $Y_{ij}^n$.  Also let $\cE_r $
be the event that message $w_r$ has a codeword typical on every
channel.
\begin{align*}
\Pr(\cE_{rj}) &= \Pr(\exists l \in \{1,2,\ldots, Q_j\}: (U^n_{j
l}(w_r),Y^n_{ij}) \in T(U_j,Y_{ij})),\quad j\in \cS\\
&\le \sum_{l=1}^{Q_j} \Pr((U^n_{j l}(w_r),Y^n_{ij}) \in
T(U_j,Y_{ij}))\\
&\le \sum_{l=1}^{Q_j} 2^{-n(I(U_j;Y_{ij})-3\delta)}\\
&\le2^{-n(I(U_j;Y_{ij})-I(U_j;Y_{ej})-3\delta +\eps_F)},
\end{align*}
where the last inequality follows since $Q_j =
2^{n(I(U_j;Y_{ej})-\eps_F)}$. Finally, the probability of $\cE_r$
can be computed as
\begin{align}
\Pr(\cE_{r}) &= \Pr(\bigcap_{j\in \cS_i} \cE_{rj}) \nonumber\\
&= \prod_{j\in \cS_i} \Pr( \cE_{rj}) \label{eq:Com_Achiev_R_Indep}\\
&= 2^{-n\sum_{j\in \cS_i}(I(U_j;Y_{ij})-I(U_j;Y_{ej})-3\eps
+\eps_F)}\nonumber\\
&= 2^{-n\sum_{j=1}^M(\{I(U_j;Y_{ij})-I(U_j;Y_{ej})\}^+-3\eps
+\eps_F)},\nonumber
\end{align}
where~\eqref{eq:Com_Achiev_R_Indep} follows by independence of
codebooks and channels. The probability of false accept event
$\cE_F$ is then given by
\begin{align*}
\Pr(\cE_F) &= \Pr(\bigcup_{r=2}^{2^{nR}} \cE_{r}) \\
&\le 2^{nR}
2^{-n\sum_{j=1}^M(\{I(U_j;Y_{ij})-I(U_j;Y_{ej})\}^+-3\delta
+\eps_F)}\\
&\le 2^{-n(3M\delta -M\eps_F +\eps_R)},
\end{align*}
which vanishes with increasing $n$ for any  $\eps_R$ and $\eps_F$
that satisfy the relation $\eps_R  > M\eps_F - 3M\delta > 0$.  The
probability of error averaged over the ensemble of codebooks is less
than \\ $\eps' = \max\left(M\delta, 2^{-n(3M\delta -M\eps_F
+\eps_R)}\right)$. This demonstrates the existence of a codebook
with error probability less than $\eps'$.

\end{itemize}

\subsubsection{ Secrecy Analysis}
We now bound the equivocation at the eavesdropper for a typical code
in the ensemble. Informally, since the codebook $\cC_j$ has
$2^{n(I(U_j;Y_{ej})-\eps_F)}$ codewords per bin, the eavesdropper's
equivocation is near perfect when observing the output of channel
$j$, i.e., $\frac{1}{n}I(W;Y_{ej}^n) \le \eps'_F$ for some $\eps'_F$
(to be specified) such that $\eps'_F \rightarrow 0$ as $\eps_F \rightarrow 0$. Since
we are sending the same message on each of the $M$ channels, the
eavesdropper can potentially reduce the equivocation by combining
the channel outputs. However in doing so, his equivocation reduces
by at most $M\eps'_F$ since the codewords on each channel are
independently selected.\footnote{It is important that the codewords
be independently selected. If they are not, say the same codeword is
repeated on each channel, the eavesdropper equivocation can be
significantly reduced by combining the channel outputs.}

The following Lemma is proved in
Appendix~\ref{app:PerfectEquivLemma1}.
\begin{lemma}
A typical code from the ensemble in our achievability scheme
satisfies the following: For any $j=1,2,\ldots M$, we have
$\frac{1}{n}I(W;Y_{ej}^n) \le \eps'_F$, where $\eps_F'=
\eps'_F(\delta,\eps_F)$ tends to zero as $\delta \rightarrow 0$ and
$\eps_F \rightarrow 0$. \label{lem:PerfectEquivLemma1}
\end{lemma}
Using the above lemma we now upper bound the mutual information at
the eavesdropper as
\begin{align}
\frac{1}{n}I(W;Y_{e1}^n,\ldots,Y_{eM}^n) &=
h(Y_{e1}^n,\ldots,Y_{eM}^n) - h(Y_{e1}^n,\ldots,Y_{eM}^n|W) \\
&= h(Y_{e1}^n,\ldots,Y_{eM}^n)- \sum_{j=1}^m h(Y_{ej}^n|W)\\
&\le \sum_{j=1}^M I(W;Y_{ej}^n) \le Mn\eps'_F,
\end{align}
where $h(Y_{e1}^n,\ldots,Y_{eM}^n|W) = \sum_{j=1}^m h(Y_{ej}^n|W)$
since the codewords in the sets $\cC_1(W), \cC_2(W),\ldots,
\cC_M(W)$ are independently selected.

Hence the normalized mutual information increases only by a fixed
amount due to observations on multiple channels. By choosing
$\eps$ in~\eqref{eq:SecrecyConstCommonMsg} to equal $M\eps'_F$, we
satisfy the secrecy constraint.

\subsection{Capacity Result of Theorem ~\ref{thm:CommonMsgCap}}
\label{subsec:proofThm1}

The result of Theorem~\ref{thm:CommonMsgCap} follows directly from
Lemma~\ref{lem:CommonMessageUpperBound}
and~\ref{lem:CommonMessageLowerBound}. For the reversely degraded
broadcast channel we have, for all $i$ and $j$, that either
$X_j\rightarrow Y_{ij}\rightarrow Y_{ej}$ or $X_j\rightarrow
Y_{ej}\rightarrow Y_{ij}$ holds. If $X_j\rightarrow
Y_{ij}\rightarrow Y_{ej}$  holds, then
\begin{align*}
I(X_j;Y_{ij}) - I(X_j;Y_{ej}) &= I(X_j;Y_{ij},Y_{ej}) -
I(X_j;Y_{ej})\\
&= I(X_j;Y_{ij}|Y_{ej}).
\end{align*}
Instead, if $X_j\rightarrow Y_{ej}\rightarrow Y_{ij}$, then
$I(X_j;Y_{ij}|Y_{ej})=0$. In either case we can write
$I(X_j;Y_{ij}|Y_{ej}) = \{I(X_j;Y_{ij}) - I(X_j;Y_{ej})\}^+$.
Substituting this in~\eqref{eq:CommonMessageUpperBound} we have
\begin{equation}
R^{+,\mrm{common}}_{K,M} \le \max_{p(X_1)p(X_2)\ldots p(X_M)}
\min_{i \in \{1,2,\ldots, K\}}
\sum_{j=1}^M\{I(X_j;Y_{ij})-I(X_j;Y_{ej})\}^+,
\label{eq:common_message_secrecy_RD}
\end{equation}
which coincides with our achievable rate
in~\eqref{eq:CommonMessageLowerBound} when we choose $U_j=X_j$.

%We note that the result also holds when the users on each channel
%are stochastically degraded, we note that by definition the set
%$\cP_j$ in~\eqref{eq:CommonMessageUpperBound} includes the joint
%distribution when the users are physically degraded.
%Hence~\eqref{eq:common_message_secrecy_RD} still holds as an upper
%bound and the capacity result follows.

As a special case of Theorem~\ref{thm:CommonMsgCap}, we have the
following corollary for the case of one receiver and one
eavesdropper.
\begin{corol}[Single User case]
Consider the reversely degraded parallel channels in
Definition~\ref{defn:reverselyDegradedBC} with one receiver and
one eavesdropper. The secrecy capacity is given by
\begin{equation}
C_{1,M} = \max_{p(X_1)p(X_2)\ldots p(X_M)} \sum_{j=1}^M
I(X_j;Y_j|Y_{ej})\;. \label{eq:SingleUserCapacity}
\end{equation}
\label{corol:SingleUserCap}
\end{corol}

\begin{remark}
The single user result admits a simple coding scheme. Split the
message $W$ into $M$ sub-messages $W_1,W_2,\ldots, W_M$ and
independently encode and decode message $W_j$ on channel $j$ with a
codebook of rate $R_j = I(X_j;Y_{j}|Y_{ej})$. With multiple
receivers however, this simple scheme is limited by the worst user on each parallel
channel and does not achieve the secrecy capacity.
\end{remark}

\subsection{Sub-optimality of a Single Codebook scheme}

The capacity of common message for reversely degraded broadcast
channels in Definition~\ref{defn:reverselyDegradedBC} without the
secrecy constraint is~\cite{elGamal:80}
\begin{equation}C^\mrm{No~Secrecy}_{K,M} =
\max_{\prod_{j=1}^M p(X_j)} \min_{i \in \{1,2,\ldots, K\}}
\sum_{j=1}^M I(X_j;Y_{ij}).~\label{eq:CommonMsgNoSec}
\end{equation}

The achievability scheme in \eqref{eq:CommonMsgNoSec} uses a
\emph{single} codebook with codewords of dimension $M \times n$.
The $j^{th}$ component of the codeword is a length $n$ sequence
sampled from an i.i.d.\ $p(X_j)$ distribution and is transmitted
on channel $j$.

\begin{figure}
\centering
\includegraphics[scale=0.5]{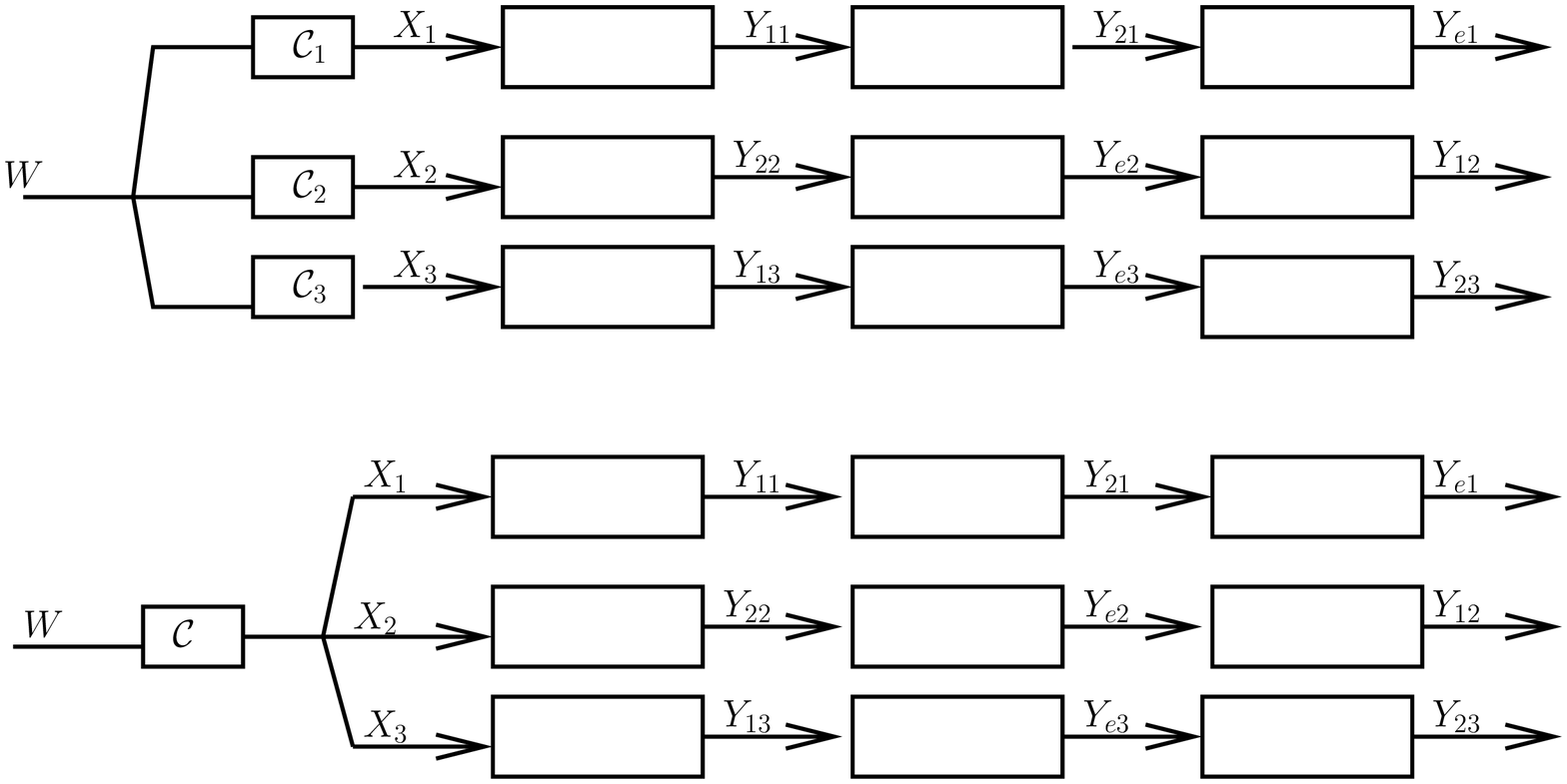}
 \caption{Two coding schemes for common message transmission on proposed
channels. The top figure shows the scheme proposed in
Theorem~\ref{thm:CommonMsgCap}. It achieves the common message
capacity. In this scheme we use independent codebooks on each
parallel channel. This allows us to separately bin on each
channel. The lower figure shows the scheme that uses a single
codebook. While this scheme is optimal when there is no
eavesdropper~\cite{elGamal:80}, it is suboptimal in our setup.
This drawback of this scheme is that because of the single
codebook, one cannot separately bin for each channel. }
\label{fig:multicastScheme}
\end{figure}

Our achievable scheme is different from this single codebook
scheme since we use independent codebooks on each parallel
channel. Note that this distinction is important in achieving
the secrecy capacity in Theorem~\ref{thm:CommonMsgCap}. The
distinction between these schemes is shown in
Fig.~\ref{fig:multicastScheme}. An achievable rate using the
single-codebook scheme in our setup is
\begin{equation}
R^{\mrm{single}} = \max_{p(X_1,X_2,\ldots, X_M)} \min_{i \in
\{1,2,\ldots K\}} \left\{{I(X_1,X_2\ldots X_K ; Y_{i1}, \ldots
Y_{iK}) -I(X_1,X_2\ldots X_K ; Y_{e1}, \ldots Y_{eK}) }\right\}.
\label{eq:CommonMsgJoint}
\end{equation}

Note that, in general, the rate~\eqref{eq:CommonMsgJoint} is
smaller than~\eqref{eq:CommonMsgCap}.\footnote{The two expressions
coincide if, for example, the eavesdropper is degraded with respect
to all the receivers on all the channels, i.e., $X_j \rightarrow
Y_{ij} \rightarrow Y_{ej}$ for every $1 \le i \le K$ and $1 \le j
\le M$.} The intuition behind this is that, by using an independent
codebook on each parallel channel, it is possibly to
\emph{separately} tune the bin size on each channel according to the
degradation of the eavesdropper. Finally note that our proposed
scheme also provides an alternative way to \cite{elGamal:80} to
achieve the common message capacity in the absence of an
eavesdropper.

\subsection{Gaussian Channels}
\label{subsec:GaussianChannelsCommonMessage}

We consider the Gaussian channel model where
\begin{equation}
\begin{aligned}
Y_{ij} &= X_{j} + Z_{ij} \\
Y_{ej} &= X_{j}+Z_{ej},
\end{aligned}
\label{eq:GaussCommonMsgModel}
\end{equation}
with $Z_{ij} \sim \cN(0,\sigma_{ij}^2)$ and $Z_{ej}\sim
\cN(0,\sigma_{ej}^2)$. All these noise variables are assumed
independent. We also impose an average power constraint
$E[\sum_{j=1}^M X_j^2]\le P$.
\begin{corol}
The common-message-secrecy-capacity for the Gaussian parallel
broadcast channel in~\eqref{eq:GaussCommonMsgModel} is
\begin{equation}
C^{\mrm{common},Gaussian}_{K,M} = \max_{(P_1,P_2,\ldots P_M)\in
\cF} \min_{1\le i \le K} \sum_{j=1}^M
\left\{\frac{1}{2}\log\left(1 + \frac{P_j}{\sigma_{ij}^2}\right) -
\frac{1}{2}\log\left(1 +
\frac{P_j}{\sigma_{ej}^2}\right)\right\}^+,
\label{eq:GaussianCommonMessageCap}
\end{equation}
where $\cF$ is the set of all feasible power allocations that
satisfy $\sum_{j=1}^M P_j \le P$.
\label{corol:GaussianCommonMessageCap}
\end{corol}
%\begin{remark}
%The optimization problem in~\eqref{eq:GaussianCommonMessageCap} is
%a convex optimization problem, however an analytical expression
%for the power allocation seems to give little insight.
%\end{remark}

To prove Corollary~\ref{corol:GaussianCommonMessageCap}, first
observe that the channel in~\eqref{eq:GaussCommonMsgModel} has the
same capacity as the corresponding reversely degraded broadcast
channel (see Fact~\ref{fact:SATO}) given by the following model: on
channel $j$, let $\pi_j(1),\ldots, \pi_j(K+1)$ denote set of
intended receivers and eavesdropper ordered from the strongest to
the weakest. For each $0\le k\le K$, the channel for user
$\pi_j(k+1)$ is $\hat{Y}_{\pi_j(k+1)j} = \hat{Y}_{\pi_j(k)j} +
\hat{Z}_{kj}$ with $Y_{\pi_j(0)j}\defeq X_j$ and
${\sigma}^2_{\pi(0)j}\defeq0$. The noise random variables
$\displaystyle \hat{Z}_{kj} \sim \cN(0,{\sigma}^2_{\pi(k+1)j}-
{\sigma}^2_{\pi(k)j})$ are independent.

Since $I(X_j;\hat{Y}_{ij}|\hat{Y}_{ej})$ is a continuous and concave
function in $p(X)$ (see Fact~\ref{fact:concavity}), we use
discretization arguments (see, e.g., Ch. 7 in~\cite{gallager}) to
extend Theorem~\ref{thm:CommonMsgCap} to the Gaussian case
\begin{equation}
C^\mrm{common}_{K,M}(P) = \max_{\substack{\prod_{j=1}^M p(X_j),\\
E[\sum_{j=1}^M X_j^2]\le P}} \min_{i \in \{1,2,\ldots, K\}}
\sum_{j=1}^M I(X_j;\hat{Y}_{ij}|\hat{Y}_{ej})\;.
\label{eq:CommonMsgCapPC}
\end{equation}
Now observe that $\max_{p(X_j), E[X_j^2]\le
P_j}I(X_j;\hat{Y}_{ij}|\hat{Y}_{ej})$ denotes the capacity of a
Gaussian wiretap channel~\cite{leung-Yan-CheongHellman:99}.
Accordingly we have
\begin{equation}
\max_{p(X_j), E[X_j^2]\le P_j}I(X_j;\hat{Y}_{ij}|\hat{Y}_{ej}) =
\left\{\frac{1}{2}\log\left(1 + \frac{P_j}{\sigma_{ij}^2}\right) -
\frac{1}{2}\log\left(1 +
\frac{P_j}{\sigma_{ej}^2}\right)\right\}^+\;.
\label{eq:Common_message_gaussian_wiretap}
\end{equation}
One then deduces~\eqref{eq:GaussianCommonMessageCap}.

\section{Parallel Channels: Independent Messages}
\label{subsec:IndepMsgParallel} We consider the case of $M$
parallel channels, one eavesdropper and $K$ receivers, each
interested in an independent message. Each such message must be
protected from the eavesdropper. We now define the achievable rate
for the case of independent messages.

%\subsection{Definition of Achievable Rate Tuple}

\begin{defn}[Length $n$ Code]
\label{def:code} A $(2^{nR_1},2^{nR_2},\ldots, 2^{nR_K},n)$ code for
the product broadcast wiretap channel in
Definition~\ref{defn:ProductBroadcastChannels} consists of a mapping
$\omega_n : \cW_1 \times \cW_2 \times \ldots \times \cW_K
\rightarrow \underbrace{\cX^n \times \cX^n \ldots \cX^n}_{M \mrm{
times}}$ from the messages of the $K$ users to the $M$ channel
inputs and $K$ decoding functions $\phi_{i,n}: \underbrace{\cY^n
\times \cY^n \times \ldots \times \cY^n}_{M \text{
times}}\rightarrow \cW_i$, one at each intended receiver. We denote
the message estimate at decoder $i$ by $\hat{W_i}$. A
perfect-secrecy-rate tuple $(R_1,R_2,\ldots, R_K)$ is achievable if,
for every $\eps>0$, there is a length $n$ code such that $\Pr(W_i
\neq \hat{W}_i)\le \eps$  for all $i=1,2,\ldots, K$, and such that
the following condition is satisfied
\begin{equation}\frac{1}{n}H(W_i | W_1,\ldots, W_{i-1},
W_{i+1},\ldots W_K, Y_{e1}^n,\ldots Y_{eM}^n) \ge \frac{1}{n}H(W) -
\eps, \quad\quad i=1,2,\ldots M. \label{eq:equivocationRatePair}
\end{equation}
The secrecy-sum-capacity $C^\mrm{sum}_{K,M}$ is the supremum of
$R_1 +R_2+\ldots + R_K$ over the achievable rate tuples
$(R_1,R_2,\ldots,R_K)$.
\end{defn}

\begin{remark}
Our constraint~\eqref{eq:equivocationRatePair} provides perfect
equivocation for each message, even if all the other messages are
revealed to the eavesdropper. It may be possible to increase the
secrecy rate by exploiting the fact that the eavesdropper does not
have access to other messages. This is not considered in the present
paper.
\end{remark}

\subsection{Main Results}

Our main result is an expression for the secrecy-sum-capacity for
the reversely degraded broadcast channel in
Definition~\ref{defn:reverselyDegradedBC}.
\begin{thm}
\label{thm:SumCap} Let $\pi_j$ denote the strongest user on
channel $j$. The secrecy-sum-capacity for the reversely broadcast
channel is given by
\begin{equation}
\label{eq:SumCap} C^\mrm{sum}_{K,M} = \max_{p(X_1)p(X_2)\ldots
p(X_M)} \sum_{j=1}^M I(X_j;Y_{\pi_j} | Y_{ej}).
\end{equation}
Furthermore, the expression in~\eqref{eq:SumCap} is an upper bound
on the secrecy-sum-capacity  when only the intended users are
reversely degraded --- but the set of receivers together with the
eavesdropper is not degraded.
\end{thm}

The remainder of this section will be devoted to the proof of
Theorem~\ref{thm:SumCap} and some discussion.

%We will first show the upper bound in
%Corollary~\ref{corol:ArbitraryEavesdropper}. This will then
%immediately yield the converse for Theorem~\ref{thm:SumCap}.

%The achievability scheme is Theorem~\ref{thm:SumCap} is rather
%straightforward. We transmit to the strongest receiver on each
%parallel channel and use independent codebooks across the
%channels. It will be only briefly discussed.

\subsection{Proof of Upper Bound in Theorem~\ref{thm:SumCap}}

\label{susec:sumCapParChanUB} We establish the upper bound in
Theorem~\ref{thm:SumCap}. Suppose a genie provides the output of
the strongest receiver, $\pi_j$, to all other receivers on each
channel, i.e., on channel $j$ the output $Y_{\pi_j}^n$ is made
available to all the receivers. Because of degradation, we may
assume, without loss of generality, that each receiver only
observes $(Y_{\pi_1}^n,\ldots, Y_{\pi_M}^n)$. Clearly, such a
genie aided channel can only have a sum capacity larger than the
original channel. Since all receivers are identical, to compute
the sum capacity it suffices to consider the situation with one
sender, one receiver, and one eavesdropper.

%\begin{defn}[Genie aided channel]
%Consider $M$  independent parallel channels with one receiver and
%one eavesdropper. On channel $j$ the transition probability of
%receiver $j$ is $p(Y_{\pi_j}|X_j)$ while that of the eavesdropper
%is $p(Y_{ej}|X_j)$. \label{def:GenieAidedOneUser}
%\end{defn}

\begin{lemma}
The secrecy-sum-capacity in Theorem~\ref{thm:SumCap} is upper
bounded by the secrecy capacity of the genie aided channel, i.e.,
$C^{\mrm{sum}}_{K,M}\le C^\mrm{Genie Aided}$.
\label{lem:GenieAidedChannel}
\end{lemma}

\begin{proof}
Suppose that a secrecy rate point $(R_1,R_2,\ldots R_K)$ is
achievable for the $K$ user channel in Theorem~\ref{thm:SumCap} and
let the messages be denoted as $(W_1,W_2,\ldots W_K)$. This implies
that, for any $\eps>0$ and $n$ large enough, there is a length $n$
code such that $\Pr(\hat{W}_i \neq W_i) \le \eps$ for
$i=1,2,\ldots,K$, and such that
\begin{equation}
\frac{1}{n}H(W_i | W_1,\ldots W_{i-1},W_{i+1},\ldots W_K,
Y_{e1}^n, Y_{e2}^n,\ldots,Y_{eM}^n) \ge R_i -\eps\;.
\label{eq:EquivCondnParChan}
\end{equation}

We now show that a rate of $(\sum_{i=1}^K
R_i,\underbrace{0,\ldots,0}_{K-1})$ is achievable on the genie
aided channel. First, note that any message that is correctly
decoded on the original channel is also correctly decoded by user
$1$ on the genie aided channel. It remains to bound the
equivocation on the genie aided channel when the message to
receiver $1$ is $W = (W_1,W_2,\ldots, W_K)$. We have
\begin{align*}
\frac{1}{n}H(W| Y_{e1}^n, Y_{e2}^n,\ldots,Y_{eM}^n) &=
\frac{1}{n}H(W_1, W_2,\ldots, W_K | Y_{e1}^n,
Y_{e2}^n,\ldots,Y_{eM}^n)\\
 &\ge \sum_{i=1}^K \frac{1}{n}H(W_i |
W_1,\ldots W_{i-1},W_{i+1},\ldots W_K, Y_{e1}^n,
Y_{e2}^n,\ldots,Y_{eM}^n)\\
&\ge \sum_{i=1}^K R_i - K\eps
\end{align*}
where the last step follows from~\eqref{eq:EquivCondnParChan}.
Since $\eps$ is arbitrary, this establishes the claim.
\end{proof}

\begin{lemma}
The secrecy capacity of the genie aided channel is
\begin{equation}
C^\mrm{Genie Aided}= \max_{p(X_1)p(X_2)\ldots p(X_M)}
\sum_{j=1}^M I(X_j;Y_{\pi_j}|Y_{ej}).
\label{eq:GenieAidedUpperBoundCap}
\end{equation}
\end{lemma}
\begin{proof}
Since all receivers are identical on the genie aided channel, this
Lemma is a direct consequence of
Corollary~\ref{corol:SingleUserCap} when specialized to the
case of $K=1$ receiver.
\end{proof}

\begin{remark}
 The upper bound continues to hold even if
the eavesdroppers channel is not ordered with respect to the
intended receivers. In general, following 
Lemma~\ref{lem:CommonMessageUpperBound}, the upper bound can be
tightened by considering, for all $1\le j \le M$,  the worst joint
distribution $p'(Y_{\pi_j},Y_{ej}|X_j)$ among all joint
distributions with the same marginal distribution as
$p(Y_{\pi_j}|X_j)$ and $p(Y_{ej}|X_j)$, yielding
\begin{equation}
\label{eq:SumCapSato} C^\mrm{sum}_{K,M} \le \min_{\prod_{j=1}^M p'(Y_{\pi_j},
Y_{ej}|X_j)}\max_{\prod_{j=1}^M p(X_j)} \sum_{j=1}^M
I(X_j;Y_{\pi_j} | Y_{ej}).
\end{equation}

\end{remark}

\subsection{Achievability Scheme}

Our achievability scheme for Theorem~\ref{thm:SumCap} requires the
receivers and the eavesdropper to be reversely degraded. We only
send information intended to the strongest user, i.e., only user
$\pi_j$ on channel $j$ can decode. It follows from the result of the
wiretap channel~\cite{wyner:75Wiretap} that a rate of $R_j =
\max_{p(X_j)}I(X_j;Y_{\pi_j}|Y_{e_j})$ is achievable on channel $j$.
Accordingly the total sum rate of $\sum_{j}R_j$ is achievable which
is the capacity expression.

\begin{remark}
The ``opportunistic transmission" strategy in
Theorem~\ref{thm:SumCap} has been previously observed in the absence
of an eavesdropper~\cite{Tse:97,liGoldsmith:98} in the context of
fading channels. Hence our result states that the optimality of
opportunistic transmission also holds in the presence of an
eavesdropper. Our converse technique, when applied to the case of no
eavesdropper, also provides a simpler argument for the optimality of
opportunistic transmission studied in ~\cite{Tse:97,liGoldsmith:98}.
%without resorting to the Kuhn-Tucker conditions and optimality of
%Gaussian codebooks~\cite{elGamal:80}.
\end{remark}

\subsection{Gaussian Channels}

Theorem~\ref{thm:SumCap} can be extended to the case of Gaussian parallel
channels. Let
$\sigma^2_{\pi_j}$ denote the noise variance of the strongest user
on channel $j$. Then the secrecy-sum-capacity is given by
\begin{equation}
C^\mrm{sum,Gaussian}_{K,M}(P) = \max_{(P_1,P_2,\ldots P_M)}
\sum_{j=1}^M \left\{\frac{1}{2}\log\left(1 +
\frac{P_j}{\sigma_{\pi_j}^2}\right) - \frac{1}{2}\log\left(1 +
\frac{P_j}{\sigma_{ej}^2}\right)\right\}^+ \label{eq:SumGaussian}
\end{equation}
where the maximization is over all power allocations satisfying
$\sum_{j=1}^M P_j\le P$. The achievability follows by using
independent Gaussian wiretap codebooks on each channel and only
considering the strongest user on each channel. For the upper
bound we have to show that Gaussian inputs are optimal in the
capacity expression in Theorem~\ref{thm:SumCap}. The
justifications are the same as in the common message case in
Section~\ref{subsec:GaussianChannelsCommonMessage}.

\section{Fading Channels}
\label{sec:fading_chan}

The second part of this paper considers wireless fading channels.
The case of one receiver and one eavesdropper has been recently
studied in a number of recent
works~\cite{barros:06,liangPoor:06a,liYatesTrappe:06a,NegiGoel05,khistiTchamWornell:06,gopalaLaiElGamal:06Secrecy}.
The proposed schemes adapt the transmission power and/or rate
depending on the instantaneous channel conditions. The time varying
nature of the fading channel enables secure transmission even when
the eavesdropper has an average channel stronger than that of the
intended receiver.

To the best of our knowledge, the above works do not consider secure
transmission to \emph{multiple} receivers in a wireless fading
environment. We first consider the case when a common message has to
be delivered to a set of intended receivers. Next, we consider the
case when each receiver obtains an independent message. For this
setting, we present a scheme based on multiuser diversity that
achieves the sum capacity in the limit of a large number of
receivers.

\subsection{Channel Model}
A block fading channel model for a system with one sender, $K$
receivers, and one eavesdropper is of the form
\begin{equation}
\by_i(t) = h_i(t)\bx(t) + \bz_i(t), \quad i \in \{1,2,\ldots K,
e\},\quad t \in \{1,2,\ldots n\} \label{eq:FastFadingVector}
\end{equation}
where $i$ denotes the index of the receiver and $t$ denotes the time
index. The vectors $\bz_i, \bx, \by_i$ are $T$ dimensional complex
valued vectors, where $T$ denotes the coherence period of the
channel. The channel coefficients $h_i(t)$ are constant over a block
of $T$ symbols and change independently over the blocks.

In our analysis we focus only on the fast-fading scenario, i.e.,
$T=1$. Using interleaved codebooks we can realize the fast-fading
case even when $T>1$. The fast-fading channel model is of the
form\begin{equation} y_i(t) = h_i(t)x(t) + z_i(t), \quad i \in
\{1,2,\ldots K, e\},\quad t \in \{1,2,\ldots n\}
\label{eq:FastFading}
\end{equation}
where the $h_i(t)$'s are sampled independently from $\CN(0,\mu_i)$
distribution and all the noise variables are sampled independently
according to $\CN(0,1)$. The input satisfies an average power
constraint $E[|X(t)|^2]\le P$.

%\begin{remark}
%The restriction to the fast-fading scenario does not incur penalty
%in the capacity in block fading channels, in the absence of the
%eavesdropper.
%
%In the absence of an eavesdropper, the capacity region is not
%sensitive to the precise value of $T$. Provided one can choose the
%block length to be sufficiently larger than $T$, the ergodicity of
%the channel renders the capacity to be insensitive to the value of
%$T$. Often, it is convenient to do the analysis for the case when
%$T=1$ and this case is sometimes referred to as the ``fast-fading"
%model. A practical way to realize the fast-fading effect is to
%employ an interleaved codebook architecture~\cite{tseViswanath:05}.
%
%
%In contrast for the wiretap setting the capacity may depend on the
%coherence period. We will consider the fast fading case i.e., when
%$T = 1$. Our analysis will provide a lower bound on the capacity for
%larger coherence periods, as the case $T=1$ can be realized via an
%interleaved codebook architecture. Such a reduction may be appealing
%in practice as the coherence periods can be different across the
%users. On the other hand, we do not consider the possibility of
%exploiting longer coherence periods for improving the secrecy
%capacity.
%\end{remark}
%

Throughout, we assume the $h_i(t)$'s to be revealed to the
transmitter, the $K$ intended receivers and the eavesdropper in a
causal manner. Implicitly we assume that there is an authenticated
public feedback link from the receivers to the transmitter. The
channel coefficients of the eavesdropper $\{h_e(t)\}_{1\le t\le n}$
are only known to the eavesdropper. The transmitter and the intended
receivers only have statistical knowledge of the eavesdropper's
channel gains.
\begin{remark}
In our setup we are assuming only one eavesdropper. Note however
that the equivocation term depends only on the statistics of
$H_e(t)$ and not on the realization of $h_e(t)$. Accordingly the
number of eavesdroppers does not matter as long as they are
statistically equivalent and do not collude.
\end{remark}

\section{Fading Channels: Common Message}

Secure multicasting refers to the case when each receiver is only
interested in a common message. The transmitter exploits the channel
knowledge of intended receivers to selectively broadcast the message
to these receivers, while the eavesdropper remains ignorant of the
message. Note that without the secrecy constraint, a non adaptive
scheme such as the one that does a flat power allocation with no
transmitter CSI, appears to be not too far from the optimal. In
contrast such schemes reveal the message to an eavesdropper with a
channel statistically equivalent to some intended receiver.

Perhaps, an interesting question is the scaling of the secrecy
capacity with the number of receivers. Does the capacity decay to
zero with the number of receivers? Note that the scheme that
consists in sending information only when all the users have a
strong channel performs poorly. Since the channel gains across the
users are independent, the achievable rate decays to zero
exponentially in the number of users.

%How can the transmitter CSI be taken into account efficiently? One
%(naive) scheme is to use an on-off transmission policy: transmit
%only when all the users simultaneously have the channel gain above a
%threshold. Unfortunately, such a scheme does not scale well with the
%number of receivers. In order to do better we must transmit
%information when only a subset of receivers are stronger than the
%threshold. However such schemes will need to schedule the
%transmissions so that all the receivers are able to eventually
%decode the message. The problem of multicasting is interesting
%because of the difference in channel gains across the receivers

An obvious upper bound is the secrecy capacity with a single
receiver. Accordingly, the best we hope for is that the common
message secrecy-capacity is a constant, independent of the number of
intended receivers. In what follows, we present a coding scheme
whose achievable rate is also a constant, independent of the number of
intended receivers. While our proposed scheme provides optimal
scaling, the precise value of the constant remains an open problem.

We now provide a formal definition of the common-message-secrecy-capacity.
\begin{defn}
\label{def:Wireless_code} A $(n,2^{nR})$ code for the channel
consists of an encoding function which is a mapping from the message
$w \in\{1,2,\ldots, 2^{nR}\}$ to transmitted symbols $x(t) =
\omega_t(w; h_1^t, h_2^t, \ldots, h_K^t)$ for $t=1,2,\ldots,n$, and
a decoding function at each receiver $\hat{W_i} = \phi_i(y_i^n;
h_1^n, h_2^n, \ldots, h_K^n)$ for each $i = 1,2, \ldots, K$. A rate
$R$  is achievable if, for every $\eps>0$, there exists a length $n$
code such that $\Pr(\hat{W}_i \neq W)\le \eps$ for $i = 1,2, \ldots,
K$ and such that
\begin{equation}
\frac{1}{n}H \left(W \mid H_e^n, H_1^n,H_2^n,\ldots,H_K^n\right) \ge
R - \eps. \label{eq:Wireless_EquivCondn_CommonMsg}
\end{equation}
\end{defn}

The entropy term in ~\eqref{eq:Wireless_EquivCondn_CommonMsg} is
conditioned on $H_1^n,\ldots, H_K^n$ as the channel gains of the $K$
receivers are assumed to be known to the eavesdropper.

\subsection{Main Results}

Our main result is an achievable rate for the common-message-secrecy-rate to $K$ receivers.
\begin{thm}
\label{thm:Wireless_CommonMsg} An achievable
common-message-secrecy-rate for the channel
model~\eqref{eq:FastFading} is given by
\begin{equation}
R^\mrm{common}(P) = \min_{1\le i \le K}E\left[\{\log(1+ |H_i|^2 P) -
E[\log(1 + |H_e|^2 P)]\}^+\right] \label{eq:Wireless_Common_Msg_Cap}
\end{equation}
\end{thm}

If all the users are i.i.d.\ Rayleigh faded with $E[|H_i|^2]=1$, the
following can be readily verified
\begin{equation}
\lim_{P\rightarrow \infty} R^\mrm{common}(P) = 0.7089
~\mrm{bits/symbol} \label{eq:Wireless_Common_Msg_HighSNR}
\end{equation}

Note that the achievable rate in~\eqref{eq:Wireless_Common_Msg_Cap}
and ~\eqref{eq:Wireless_Common_Msg_HighSNR} does not depend on the
number of receivers. Accordingly, we do not subscript the rate by
$K$. That the capacity does not decay with the number of receivers
is the best scaling of the capacity with the number of receivers
that one can expect.

The ``interesting" part of our achievability
rate~\eqref{eq:Wireless_Common_Msg_Cap} is the $\{\cdot\}^+$ inside
the expectation. This is essentially a consequence of the multiple
codebook scheme we presented for the parallel channel case in
Section~\ref{subsec:CommonMsgParallel}.

Our approach to establish the achievability
of~\eqref{eq:Wireless_Common_Msg_Cap} is to decompose the fading
channel into a set of parallel channels and invoke
Lemma~\ref{lem:CommonMessageLowerBound} for the \emph{probabilistic}
extension of the parallel broadcast channel.

\subsection{Achievability Scheme}
\label{subsec:CommonMsgSchemeI} First we consider the following
\emph{probabilistic extension} of the parallel broadcast
channel~\cite{liGoldsmith:98}: At each time, only one of the
parallel channel operates and channel $j$ is selected with a
probability $p_j$, independent of all other times. Also suppose that
there is a total power constraint $P$ on the input. A
straightforward extension of Lemma~\ref{lem:CommonMessageLowerBound}
provides the following achievable rate \begin{equation}
R^{\mrm{common}}_{K,M}(P)
\defeq \max \min_{
i \in \{1,2,\ldots,K\}} \sum_{j=1}^M p_j\{I(U_j;Y_{ij}) - I(U_j;
Y_{ej})\}^+, \label{eq:ProbabilisticCommonMessageLowerBound}
\end{equation}
where $U_1,U_2,\ldots U_M$ are auxiliary random variables and the
maximum is over the product distribution ${p(U_1)p(U_2)\ldots
p(U_M)}$ and the stochastic mappings  $X_j = f_j(U_j)$ that satisfy
$ \sum_{j=1}^M p_jE[X_j^2]\le P$.
%\label{lem:ProbabilisticCommonMessageLowerBound}
%\end{lemma}

%The basic idea in our achievability scheme is to use the result of
%parallel channels in
%Lemma~\ref{lem:ProbabilisticCommonMessageLowerBound}.

Next, we map the fading channel~\eqref{eq:FastFading} into a set of
parallel channels and invoke the achievable
rate~\eqref{eq:ProbabilisticCommonMessageLowerBound}. However, we
need to resolve the technicality in that the fading channel has
continuous valued fading coefficients, while the rate expression
in~\eqref{eq:ProbabilisticCommonMessageLowerBound} is only for a
finite number of parallel channels.
Following~\cite{goldsmithVaraiya:97}, our approach is to
\emph{discretize} the continuous valued coefficients and thus create
parallel channels, one for each quantized state. The number of
parallel channels increases as the quantization becomes finer. In
what follows we only quantize the magnitude of the fading
coefficients. The receiver can always rotate the phase, so it plays
no part.

We quantize the channel gains into one of the $q$ values: $A_1=0<
A_2 < \ldots < A_{q+1} = \infty$. Receiver $i$ is in state
$l\in\{1,2,\ldots,q\}$ at time $t$ if $A_{l} \le |H_{i}(t)|^2 <
A_{l+1}$. When in state $l$, the receiver's channel gain is
pessimistically discretized to $\sqrt{A_l}$. Since there are $K$
independent users, there are a total of $M=q^K$ possible
super-states, which we number as $S_1, S_2,\ldots,S_M$. Denote the
quantized gain of user $i$ in $S_j$ by the double subscript
$S_{ij}$. Let $p(S_j)$ denote the probability of state $S_j$. Also
let $p_i(A_l)$ be the probability that a user $i$ is in state $l$
i.e., $p_i(A_l) = \sum_{k=1: S_{ik}= A_l}^M p(S_k)$. In super-state
$S_j$, the channel of user $i$ and the
eavesdropper are\begin{align*} y_{ij}(t) = \sqrt{S_{ij}} x(t) + z_i(t), \\
 y_{el}(t) = H_{e}(t) x(t) + z_e(t).\end{align*} By
selecting $U_j \sim \CN(0,P)$ and $X_j = U_j$, the argument in the
summation in~\eqref{eq:ProbabilisticCommonMessageLowerBound} (with
the eavesdropper output $(Y_{el},H_{e})$) is
\begin{align*}
\{I(U_j;Y_{ij}) - I(U_j; Y_{ej},H_{e}) \}^+
&= \{I(X_j;Y_{ij}) - I(X_j; Y_{ej},H_e)\}^+\\
&= \{I(X_j; \sqrt{S_{ij}}X_j + Z_i)- I(X_j; H_{e}X_j + Z_e,H_e)\}^+\\
&=\{\log(1 + S_{ij} P) - E[\log(1 + |H_e|^2 P]\}^+.
\end{align*} Substituting in~\eqref{eq:ProbabilisticCommonMessageLowerBound}, we have that the
following rate is achievable
\begin{align}
R_Q^\mrm{common}(P) &= \min_{1\le i \le K } \sum_{j=1}^M p(S_j)
\{\log(1 + S_{ij}P)-E[\log(1 + |H_e|^2P)]\}^+\\
&=\min_{1\le i \le K } \sum_{l=1}^q p_i(A_l)\{\log(1 + A_l
P)-E[\log(1 + |H_e|^2P)]\}^+,
\end{align}where the second equality follows from rewriting the
summation over the states of each individual user. As $q\rightarrow
\infty$, the above sum converges to
\begin{align}
& \min_{1\le i \le K}\int_{0}^\infty\{\log(1+ xP)
- E[\log(1 + |H_e|^2 P)]\}^+p_i(x)dx\\
&=\min_{1\le i \le K}E\left[\{\log(1+ |H_i|^2 P) - E[\log(1 +
|H_e|^2 P)]\}^+\right],
\end{align}
yielding~\eqref{eq:Wireless_Common_Msg_Cap}.

\begin{remark}
\label{remark:wirelesscommonmessageSchemeII} The scheme presented
above requires $q^K$ codebooks, where $q$ is the number of
quantization bins. A different decomposition which requires only
$2^K$ codebooks and provides the same achievable rate is presented
in Appendix~\ref{app:commonMsgSchemeII}. This scheme can be
implemented in practice with an outer erasure code and an inner
wiretap code, as discussed in~\cite{khistiTchamWornell:06}.

\end{remark}

\section{Fading Channels: Independent Messages}
We consider the case where each receiver wants an independent
message. We will only focus on the sum rate of the system. This
scenario has been widely studied in conventional systems (i.e.,
without a secrecy constraint) where the transmitter CSI provides
dramatic gains (see e.g.,~\cite{Tse:97,liGoldsmith:98}). An
``opportunistic scheme" that selects the user with the largest
instantaneous gain maximizes the sum-rate of the system. The results
in this section can be interpreted as an extension of opportunistic
transmission in the presence of eavesdroppers.

\begin{defn}
\label{def:Wireless_code2} A  $(n,2^{nR_1},\ldots,2^{nR_K})$ code
consists of an encoding function from the messages $w_1,\ldots, w_K$
with $w_i \in \{1,2,\ldots, 2^{nR_i}\}$ to transmitted symbols $x(t)
= \omega_t(w_1,w_2,\ldots,w_K; h_1^t, h_2^t, \ldots, h_K^t)$ for
$t=1,2,\ldots,n$, and a decoding function at each receiver
$\hat{W_i} = \phi_i(y_i^n; h_1^n, h_2^n, \ldots, h_K^n)$. A rate
tuple $(R_1,R_2,\ldots,R_K)$ is achievable with perfect secrecy if,
for any $\eps>0$, there exists a length $n$ code such that, for each
$i=1,2,\ldots,K$, with $W_i$ uniformly distributed over
$\{1,2,\ldots,2^{nR_i}\}$, we have $\Pr(\hat{W_i}\neq W_i)\le \eps$
and
\begin{equation}
\frac{1}{n}H \left(W_i \Biggm|
W_1,\ldots,W_{i-1},W_{i+1},\ldots,W_K, H_e^n,
H_1^n,\ldots,H_K^n\right) \ge R_i - \eps.
\label{eq:Wireless_EquivCondn}
\end{equation}
The secrecy-sum-capacity is the supremum value of $R_1  +
R_2+\ldots+ R_K$ among all achievable rate tuples.
\end{defn}

\subsection{Main Results}
%Our main result in this section is that opportunistic transmission
%in conjunction with Gaussian wiretap codebooks for each user
%achieves the sum capacity in the limit of large number of users.
%Let $|H_\mrm{max}|^2 = \max\{|H_1|^2, |H_2|^2,\ldots, |H_K|^2\}$,
%i.e.,

In the following, let $H_\mrm{max}$ denote the largest instantaneous
channel gain among the $K$ users. We first upper and lower bound the
secrecy-sum-capacity.

\begin{lemma}
For the channel model~\eqref{eq:FastFading}, the
secrecy-sum-capacity is upper and lower bounded as
\begin{equation}R^\mrm{+}_{K}(P) =
\max_{P(H_\mrm{max}): E[P(H_\mrm{max})]\le P}
E\left[\{{\log(1+|H_\mrm{max}|^2 P(H_\mrm{max}))}-{\log(1+ |H_e|^2
P(H_\mrm{max}))}\}^+\right]
\label{eq:Wireless_SumCapacityUpperBound}\end{equation}
and

%\begin{lemma}
%For the channel model~\eqref{eq:FastFading}, the
%secrecy-sum-capacity is lower bounded as follows:
\begin{equation}R^\mrm{-}_{K}(P) =\max_{P(H_\mrm{max}): E[P(H_\mrm{max})]\le P}
E\left[{\log(1+|H_\mrm{max}|^2 P(H_\mrm{max}))}-{\log(1+ |H_e|^2
P(H_\mrm{max}))}\right]\label{eq:Wireless_SumCapacityLowerBound},\end{equation}
\label{lem:Wireless_SumCapacityLowerBound}
respectively, where $\{v\}^+$ denotes the $\max(0,v)$.

\end{lemma}

The difference in our lower and upper bounds
in~\eqref{eq:Wireless_SumCapacityUpperBound}
and~\eqref{eq:Wireless_SumCapacityLowerBound} is that the
$\{\cdot\}^+$ operator is inside the expectation in our upper bound
but not in the lower bound. Thus the ``loss" with respect to the
upper bound occurs whenever $|H_\mrm{max}|^2 \le |H_e|^2$. As the
number of intended receivers grows this event happens rarely and the
gap between the upper and lower bounds vanishes. Formally we have

\begin{thm}
\label{thm:WirelessAsymptoticSumCapacity} The gap between our upper
bound $R_K^+(P)$ and the lower bound $R_K^-(P)$ in
Lemma~\ref{lem:Wireless_SumCapacityLowerBound} satisfies
\begin{equation}
R^+_K(P) - R^-_K(P)  \le \Pr(|H_e|^2\ge |H_\mrm{max}|^2)
E\left[\log\frac{|H_e|^2}{|H_\mrm{max}|^2}\Biggm | |H_e|^2\ge
|H_\mrm{max}|^2\right].
\end{equation}
The bounds coincide in the limit $K\rightarrow \infty$ when all the
channel gains are sampled from $\CN(0,1)$.
\begin{equation}
C^\mrm{sum}_{K}(P) = \max_{P(H_\mrm{max}): E[P(H_\mrm{max})]\le P}
E\left[{\log(1+|H_\mrm{max}|^2 P(H_\mrm{max}))}-{\log(1+ |H_e|^2
P(H_\mrm{max}))}\right]+o(1),\label{eq:WirelessSumCapFF}
\end{equation}
where $o(1)\rightarrow 0$ as $K\rightarrow \infty$.
\end{thm}

The result of Theorem~\ref{thm:WirelessAsymptoticSumCapacity} shows
that opportunistic transmission in conjunction with single user
Gaussian codebooks achieves the optimal sum secrecy-rate in the
limit of large number of receivers.

\begin{remark}
To the best of our knowledge, the secrecy-sum-capacity for a finite
number of receivers has not been resolved for the fast fading
model~\eqref{eq:FastFading}. When the coherence period is large
enough so that one can invoke random coding arguments in each
period, it appears possible to extend the single receiver result
in~\cite{gopalaLaiElGamal:06Secrecy} to determine the secrecy-sum-capacity for finite number of users. We elaborate this connection
later in the section \ref{discussion}.
\end{remark}

Our upper and lower bounds do not coincide for a finite number of
users. Nevertheless, the high SNR limit provides a convenient
operating regime for numerical evaluation of the bounds.

\begin{corol}
\label{corol:Wireless_High_SNR_Corol} We have
\begin{equation}
\label{eq:WirelessHighSNR}
\begin{aligned}
\lim_{P\rightarrow\infty}R_K^\mrm{+}(P) &=
E\left[\left\{\log\frac{|H_\mrm{max}|^2}{|H_e|^2}\right\}^+\right]\\
\lim_{P\rightarrow\infty}R_K^\mrm{-}(P) &= \max_{T\ge
0}\Pr(|H_\mrm{max}|^2 \ge
T)E\left[\log\frac{|H_\mrm{max}|^2}{|H_e|^2}\Biggm | |H_\mrm{max}|^2
\ge T\right].
\end{aligned}
\end{equation}
\end{corol}

\subsection{Upper Bound in Lemma~\ref{lem:Wireless_SumCapacityLowerBound}}

Our proof technique is to introduce a single user genie-aided
channel as in Section~\ref{subsec:IndepMsgParallel} and then to
upper bound this single user channel. This upper bound on the genie
aided channel is closely related to an upper bound provided
in~\cite{gopalaLaiElGamal:06Secrecy} for the slow fading channel. We
nevertheless provide a complete derivation in
Appendix~\ref{app:GenieAidedFadingChannel}.

\subsection{Achievability in Lemma~\ref{lem:Wireless_SumCapacityLowerBound}}
The achievability scheme combines opportunistic transmission and a
Gaussian wiretap code. At each time, only the message of the user
with the best instantaneous channel gain is selected for
transmission.

As in Section~\ref{subsec:CommonMsgSchemeI}, we quantize each
receiver's channel gain into $q$ levels $A_1=0 < A_2 < \ldots < A_q
\le A_{q+1} = \infty$. Since the channel gains of the $K$ users are
independent, there are a total of $M=q^K$ different super-states.
These are denoted as $S_1, S_2,\ldots, S_M$. Each of the
super-states denotes one parallel channel. Note that on each
parallel channel, the intended users have a Gaussian channel, while
the eavesdropper has a fading channel.

Our scheme transmits an independent message on each of the $M$
parallel channels.  Let $G_j \in \{A_1,A_2,\ldots, A_{q}\}$ denote
the gain of the strongest user on channel $j$. We use a Gaussian
codebook with power $P(G_j)$ on channel $j$. The achievable rate on
channel $j$ is
\begin{align*}
R_j &= I(U_j; Y_j) - I(U_j; Y_{ej}, H_{ej})\\
&= \log(1+ G_j P(G_j)) - E[\log(1 + |H_e|^2 P(G_j))],
\end{align*}
where the second equality follows from our choice of $X_j = U_j \sim
\cN(0, P(G_j))$. The overall achievable sum rate is given by
\begin{align*}
R^-_K(P) &= \sum_{j=1}^{M}\Pr(S_j) R_j \\
&= \sum_{j=1}^{M}\Pr(S_j) ( \log(1+ G_j P(G_j)) - E[\log(1 + |H_e|^2
P(G_j))])\\
&= \sum_{l=1}^{q}\Pr(A_l) ( \log(1+ A_l P(A_l)) - E[\log(1 + |H_e|^2
P(A_l))]),
\end{align*}
where the last equality follows by using the fact that $G_j \in
\{A_1,A_2,\ldots, A_{q}\}$ and rewriting the summation over these
indices. As $q\rightarrow\infty$,
\begin{equation}
\begin{aligned}
R_K^-(P) &= \int_{0}^\infty ({\log(1+a P(a))}-{E[\log(1+
|H_e|^2 P(a))]})p(a) da \\
&= E\left[{\log(1+|H_\mrm{max}|^2 P(H_\mrm{max}))}-{\log(1+ |H_e|^2
P(H_\mrm{max}))}\right],
\end{aligned}
\end{equation}
which establishes~\eqref{eq:Wireless_SumCapacityLowerBound}.

\subsection{Proof of Theorem~\ref{thm:WirelessAsymptoticSumCapacity}}

Let $P^*(H_\mrm{max})$ be the power allocation that maximizes
$R^+_K(P)$ in~\eqref{eq:Wireless_SumCapacityUpperBound}. We have
\begin{align*}
R^+_K(P) - R^-_K(P) &\le E\left[\{{\log(1+|H_\mrm{max}|^2
P^*(H_\mrm{max}))}-{\log(1+ |H_e|^2 P^*(H_\mrm{max}))}\}^+\right]
\\ &\quad \quad - E\left[{\log(1+|H_\mrm{max}|^2 P(H_\mrm{max}))}-{\log(1+ |H_e|^2
P(H_\mrm{max}))}\right]\\
&= \Pr(|H_e|^2\ge |H_\mrm{max}|^2) E\left[\log\frac{1+
|H_e|^2P^*(H_\mrm{max})}{1+ |H_\mrm{max}|^2P^*(H_\mrm{max})}\Biggm |
|H_e|^2\ge
|H_\mrm{max}|^2\right]\\
&\le\Pr(|H_e|^2\ge |H_\mrm{max}|^2)
E\left[\log\frac{|H_e|^2}{|H_\mrm{max}|^2}\Biggm | |H_e|^2\ge
|H_\mrm{max}|^2\right],\\
&\le \frac{1}{K+1} 2 \log 2
\end{align*}where the first step follows substituting the bounds in
~\ref{lem:Wireless_SumCapacityLowerBound}, the third step follows
from the fact that $\log\frac{1+ |H_e|^2a}{1+ |H_\mrm{max}|^2a}$ is
increasing in $a$ for $|H_e|^2\ge |H_\mrm{max}|^2$, and where the
last step follows from Lemma~\ref{lem:OrderedStatisticsBound}
(proved in the Appendix~\ref{app:WirelessAsymptoticSumCapacity}) and
the fact that $\Pr(|H_e|^2 \ge |H_\mrm{max}|^2)= 1/(1+K)$, since we
assumed the channel coefficients to be i.i.d.

\subsection{Proof of Corollary~\ref{corol:Wireless_High_SNR_Corol}}
The upper bound follows from the simple identity, that for every
$P\ge 0$,\begin{equation} \left\{\log\frac{1+|H_\mrm{max}|^2 P}{ 1+
|H_e|^2 P}\right\}^+ \le \left\{\log\frac{|H_\mrm{max}|^2 }{ |H_e|^2
}\right\}^+.
\end{equation}

For the lower bound, we use a two level power allocation strategy
in~\eqref{eq:Wireless_SumCapacityLowerBound}. Fix a threshold $T\ge
0$ and let \begin{equation}P(H_\mrm{max})=\left\{
                               \begin{array}{ll}
                                 P_0 \defeq \frac{P}{\Pr(|H_\mrm{max}|^2 \ge T)} & \hbox{$|H_\mrm{max}|^2 \ge T$} \\
                                 0 & \hbox{{otherwise}.}
                               \end{array}
                             \right.
\end{equation}
This choice gives an achievable rate of
$$R_K^-(P) = \Pr(|H_\mrm{max}|^2\ge T)E\left[\log\frac{1+|H_\mrm{max}|^2 P_0}{
1+ |H_e|^2 P_0}\Biggm | |H_\mrm{max}|^2 \ge T\right].$$

The argument inside the expectation is bounded by
$E[\log\frac{|H_\mrm{max}|^2}{|H_e|^2}]$ for all $P_0>0$. Hence by
the dominated convergence theorem the limit $P\rightarrow \infty$
and the expectation can be interchanged. Accordingly we have
\begin{align*}
\lim_{P\rightarrow\infty}R_K^-(P) &= \Pr(|H_\mrm{max}|^2\ge T)
E\left[\lim_{P\rightarrow\infty}\log\frac{1+|H_\mrm{max}|^2 P_0}{ 1+
|H_e|^2 P_0}\Biggm | |H_\mrm{max}|^2 \ge T\right]\\
&= \Pr(|H_\mrm{max}|^2\ge
T)E\left[\log\frac{|H_\mrm{max}|^2}{|H_e|^2}\Biggm | |H_\mrm{max}|^2
\ge T\right]
\end{align*}which gives the desired result.

\subsection{Discussion}

\label{discussion} Theorem~\ref{thm:WirelessAsymptoticSumCapacity}
guarantees an arbitrarily small gap between upper and lower bounds
on the sum-secrecy-capacity, that holds for any fixed coherence
period, provided the number of users is large enough.

In~\cite{gopalaLaiElGamal:06Secrecy} two schemes are presented
--- a {\emph {variable rate}} and a {\emph {constant rate}} --- for the case of a
single receiver in slow fading environment. Straightforward
extensions of these schemes for multiple receivers reveals the
following. The variable rate scheme achieves the our upper bound in
\eqref{eq:Wireless_SumCapacityUpperBound}, whereas the constant rate
scheme achieve our lower bound in
\eqref{eq:Wireless_SumCapacityLowerBound}. Since these two
expressions coincide as the number of receivers tends to infinity,
one deduces that the gains of variable rate schemes become
negligible in this limit.

\subsubsection*{Numerical Evaluation of the Upper and Lower Bounds}
\begin{figure}
\centering
\includegraphics[scale=0.5]{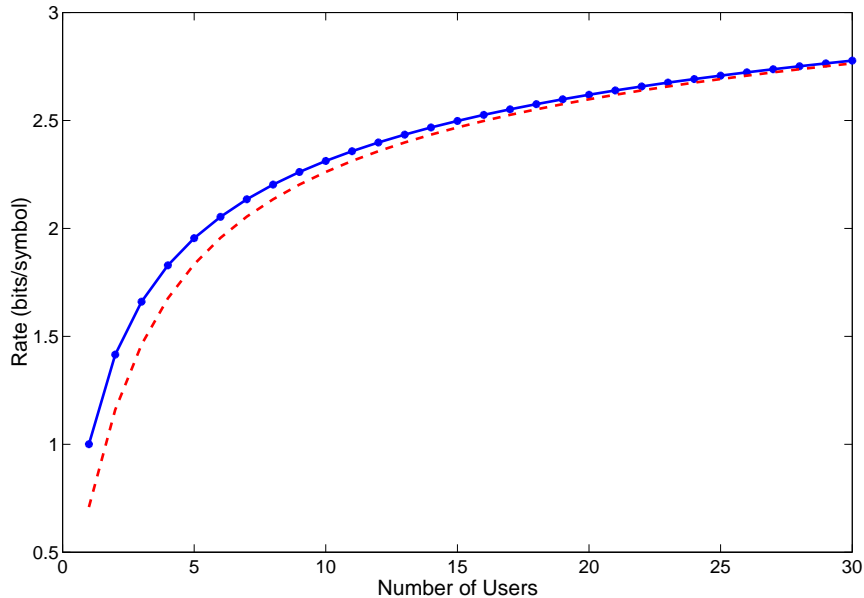}
\caption{Upper and Lower bounds in the High SNR limit
(c.f.~\eqref{eq:WirelessHighSNR}) for the i.i.d.\ Rayleigh fading
case. The y-axis plots the bounds in nats/symbol while the x-axis
plots the number of users. } \label{fig:bounds}
\end{figure}

We plot the upper and lower bounds in the high SNR limit
in~\eqref{eq:WirelessHighSNR} in Fig.~\ref{fig:bounds} for the case
of i.i.d. Rayleigh fading. Note that the bounds are quite close
even for a moderate number of users.

%\subsubsection*{Multiple Antennas}
%In conventional systems, multiple antennas have proven
%to be a valuable resource both in terms of increasing the throughput
%and in providing fairness~\cite{ViswanathTse:02}. Understanding the
%role of multiple antennas for security is a very interesting topic,
%but outside the scope of this paper.

%We briefly remark that the random beamforming based approach
%proposed for inducing fairness in multiuser diversity systems
%in~\cite{ViswanathTse:02} can be easily extended in our setting. At
%each time, the transmit antenna can randomly beamform in a certain
%direction and transmit random noise in all directions orthogonal to
%the chosen direction. Such an approach can significantly enhance the
%robustness of the system over the time diversity based approaches
%discussed in this paper.

\subsubsection*{Colluding Attacks}
We noted earlier that any number of statistically equivalent eavesdroppers does
not affect our capacity  as long as they do not collude. If the eavesdroppers collude then they can combine the received signals and attempt to decode the
message. The upper and lower bounds in
Lemma~\ref{lem:Wireless_SumCapacityLowerBound} can be extended by
replacing the term $|H_e|^2$ with $||\bH_e||^2$, where $\bH_e$ is
the vector of channel gains of the colluding eavesdroppers. One
conclusion from these bounds is that the secrecy capacity is
positive unless the colluding eavesdropper population grows as $\log
K$.

\section{Conclusion}
\label{sec:concl} 

A generalization of the wiretap channel to the case of parallel and fading
channels with multiple receivers is considered. We established the common-message-secrecy-capacity for the
case of reversely degraded parallel channels and provided upper and lower
bounds for the general case. For independent messages over parallel channels,
the sum-secrecy capacity has been determined. For fading channels, we examined
a fast fading scenario when the transmitter knows the instantaneous channels of
all the intended receivers but not of the eavesdropper. Interestingly, the common-message-secrecy-capacity does not decay to zero
as the number of intended receiver grows. For the case of independent
messages, it was shown that an opportunistic scheme achieves the
secrecy-sum-capacity in the limit of large number of users.

The protocols investigated in this paper relied on time diversity
(for the common message) and multiuser diversity (for independent
messages) to enable secure communication. In situations where such
forms of diversity is not available, it is of interest to develop a
 formulation for secure transmission, analogous to the
outage formulation for slow fading channels. 
Secondly, the impact of multiple antennas on secure transmission is far from
being clear at this stage. While multiple antennas can theoretically provide significant gains in throughput in the conventional systems, a theoretical analysis for the case of confidential messages is naturally of great interest.

%
%\begin{figure}
%\centering
%\includegraphics[width=2.5in]{myfigure}
% where an .eps filename suffix will be assumed under latex,
% and a .pdf suffix will be assumed for pdflatex
%\caption{Simulation Results}
%\label{fig_sim}
%\end{figure}

% An example of a double column floating figure using two subfigures.
% (The subfigure.sty package must be loaded for this to work.)
% The subfigure \label commands are set within each subfigure command, the
% \label for the overall fgure must come after \caption.
% \hfil must be used as a separator to get equal spacing
%
%\begin{figure*}
%\centerline{\subfigure[Case I]{\includegraphics[width=2.5in]{subfigcase1}
% where an .eps filename suffix will be assumed under latex,
% and a .pdf suffix will be assumed for pdflatex
%\hfil
%\subfigure[Case II]{\includegraphics[width=2.5in]{subfigcase2}
% where an .eps filename suffix will be assumed under latex,
% and a .pdf suffix will be assumed for pdflatex
%\label{fig_second_case}}}
%\caption{Simulation results}
%\label{fig_sim}
%\end{figure*}

\appendices

\section{Proof of Fact~\ref{fact:concavity}}
\label{app:concavityProof}

Let $T$ be a binary valued random variable such that: if $T=0$ the
induced distribution on $X$ is $p_0(X)$, i.e., $p(Y,Z,X|T=0) =
p(Y,Z|X)p_0(X)$, and if $T=1$ the induced distribution on $p(X)$ is
$p_1(X)$ i.e. $p(Y,Z,X|T=1) = p(Y,Z|X)p_1(X)$. Note the Markov chain
$T \rightarrow X \rightarrow (Y,Z)$. To establish the concavity of
$I(X;Y|Z)$ in $p(X)$ it suffices to show that
\begin{equation}
I(X;Y|Z,T) \le I(X;Y|Z).
\end{equation}
The following chain of inequalities can be verified.
\begin{align}
I(X;Y|Z,T) -I(X;Y|Z)
&= \{I(X;Y,Z|T) - I(X;Z|T)\} - \{I(X;Y,Z)-I(X;Z)\} \label{eq:concaveThm1}\\
&= \{I(X;Y,Z|T) - I(X;Z|T)\} - \{I(TX;Y,Z)-I(TX;Z)\} \label{eq:concaveThm2}\\
&= \{I(X;Y,Z|T) - I(TX;Y,Z)\} - \{I(X;Z|T)-I(TX;Z)\} \nonumber\\
&= I(T;Z) - I(T;Y,Z) = -I(T;Y|Z) \le 0 \nonumber.
\end{align}
Equation~\eqref{eq:concaveThm1} is a consequence of the chain rule
for mutual information. Equation~\eqref{eq:concaveThm2} follows from
the fact that $T\rightarrow X \rightarrow (Y,Z)$ forms a Markov
Chain, so that $I(T;Z|X)= I(T;Y,Z|X)=0$.

\section{Proof of Lemma~\ref{lem:PerfectEquivLemma1}}
\label{app:PerfectEquivLemma1} Since there are $Q_j =2^{nR_{ej}}$
codewords per message bin $\cC_j(W)$ and each codeword is equally
likely to be selected
\begin{equation}
\begin{aligned}
\frac{1}{n}H(U_j^n|W) &=  R_{ej} \\
&=  I(U_j;Y_{ej})-\eps_F,
\end{aligned}
\label{eq:Com_Achiev_Secrecy_Uniform}
\end{equation}
where the last equality follows from the definition of $R_{ej}$
in~\eqref{eq:Com_Achiev_R_Defn}.  Since the number of codewords in
each bin is less than $2^{n(I(U_j;Y_{ej})-\eps_F)}$,  we can select
a code that satisfies Fano's inequality
\begin{equation}
\frac{1}{n}H(U_j^n|W,Y_{ej}^n) \le \g\defeq\frac{1}{n}+\eps_F\;
R_{ej}\;.
 \label{eq:Com_Achiev_Secrecy_Fano}
\end{equation}
The equivocation at the eavesdropper can be lower bounded as
\begin{align}
H(W|Y_{ej}^n) &= H(W,U_j^n|Y_{ej}^n)-H(U_j^n|W,Y_{ej}^n) \nonumber\\
&\ge H(U_j^n|Y_{ej}^n) - n\g \label{eq:Com_Achiev_Secrecy_FanoA}\\
&= H(U_j^n) - I(U_j^n;Y_{ej}^n)-n\g \nonumber\\
&= H(U_j^n,W)- I(U_j^n;Y_{ej}^n)-n\g \label{eq:Com_Achiev_Secrecy_Determ}\\
&= H(W) + H(U_j^n|W)- I(U_j^n;Y_{ej}^n)-n\g \nonumber\\
&\ge H(W)+ nI(U_j;Y_{ej}) -
I(U_j^n;Y_{ej}^n)-n\g-n\eps_F.\label{eq:Com_Achiev_Secrecy_Uniform2}
\end{align}
Here~\eqref{eq:Com_Achiev_Secrecy_FanoA} follows from
substituting~\eqref{eq:Com_Achiev_Secrecy_Fano},~\eqref{eq:Com_Achiev_Secrecy_Determ}
from the fact that $W$ is deterministic given $U_j^n$
and~\eqref{eq:Com_Achiev_Secrecy_Uniform2} follows by
substituting~\eqref{eq:Com_Achiev_Secrecy_Uniform}. We now show that
for a suitably chosen $\eps'>0$
\begin{equation}
I(U_j^n;Y_{ej}^n) \le nI(U_j;Y_{ej}) + n\eps'.
\end{equation}
%Now note that $I(U_j^n;Y_{ej}^n) = H(U_j^n) - H(U_j^n | Y_{ej}^n)$.

First note the following %suppose that  $(u_j^n, y_j^n)\in
%T(U_j,Y_j)$.
\begin{equation}
\begin{aligned}
&\left| -\frac{1}{n}\log p(y_j^n)-nH(Y_j)\right| \le \delta, \quad
\forall y_j^n \in T(Y_j)\\
&\left| -\frac{1}{n}\log p(y_j^n|u_j^n)-nH(Y_j|U_j)\right| \le
\delta, \quad \forall (y_j^n,u_j^n) \in T(Y_j,U_j).
\end{aligned}
\label{eq:Com_Achiev_Secrecy_Typicality}
\end{equation}
Let $J$ be an indicator function which equals $1$ if $(y_j^n,u_j^n)
\in T(Y_j,U_j)$. From~\eqref{eq:Com_Achiev_Secrecy_Typicality} we
note that
\begin{equation}
I(U_j^n;Y_j^n|J=1) \le nI(U_j;Y_j) + 2n\delta\;.
~\label{eq:Com_Achiev_Secrecy_Typicality2}
\end{equation}

Now we can upper bound $I(U_j^n;Y_j^n)$ as
\begin{align}
I(U_j^n;Y_j^n) &\le I(U_j^n;Y_j^n,J) \nonumber\\
&= I(U_j^n;Y_j^n|J) + I(U_j^n;J) \nonumber\\
&\le I(U_j^n;Y_j^n|J=1) + I(U_j^n;Y_j^n|J=0)\Pr(J=0) + H_b(J) \label{eq:Com_Achiev_Secrecy_CondnI1}\\
&\le nI(U_j;Y_j) + 2n\delta +  n\eps \log|\cY| + 1
\label{eq:Com_Achiev_Secrecy_CondnI2},
\end{align}
where~\eqref{eq:Com_Achiev_Secrecy_CondnI1} follows from the fact
that $I(U_j^n;J) \le H_b(J)$, the binary entropy of $J$.
The inequality~\eqref{eq:Com_Achiev_Secrecy_CondnI2} follows from the fact that
$H_b(J) \le 1$, $\Pr(J=0) \le \eps$, $I(U_j^n;Y_j^n|J=0) \le n\log
|\cY|$, and~\eqref{eq:Com_Achiev_Secrecy_Typicality2}. We now select
$$\eps' = 2\delta + \delta \log |\cY|+ \frac{1}{n}.$$
Combining~\eqref{eq:Com_Achiev_Secrecy_Uniform2}
and~\eqref{eq:Com_Achiev_Secrecy_CondnI2} we have
\begin{equation}
\begin{aligned}
\frac{1}{n}I(W;Y_{ej}^n)&\le \eps' + \g + \eps_F \\
&= 2\delta + \eps |\cY|+ \frac{2}{n} + \eps_FR_{ej} + \eps_F\\
&\defeq \eps_F'
\end{aligned}
\end{equation}

\section{Alternate Scheme for Theorem~\ref{thm:Wireless_CommonMsg}}
\label{app:commonMsgSchemeII} 
\begin{figure}
\centering
\includegraphics[scale=0.5]{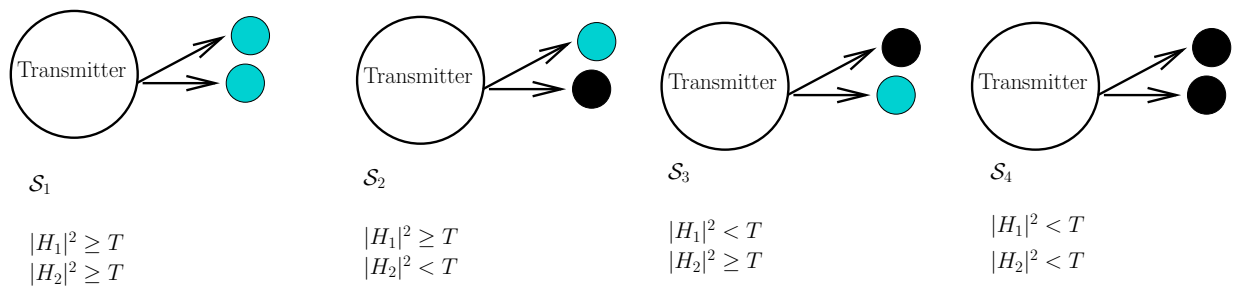}
\caption{Decomposition of the two user system into four states. In
the first state both users have channel gains above the threshold.
In the second state only user 1 has channel above the threshold
while in the third state only user 2 has channel above the
threshold. The fourth state both users have channels below the
threshold. In any state, a user is colored dark if the channel gain
is below the threshold and shaded if the channel gain is above the
threshold.} \label{fig:cmsg}
\end{figure}

We present an alternate scheme for
Theorem~\ref{thm:Wireless_CommonMsg}. For simplicity we focus on the
case of two receivers.  The case of more than two receivers
is analogous. Fix a threshold $T>0$ and decompose the
system into four states as shown in Fig.~\ref{fig:cmsg}.
. 
The transmission happens over a block of length $n$ and we classify $t = 1,2,\ldots,n$ as

\begin{equation}
\begin{aligned}
&\cS_1 = \left\{ t \in \{1,n\}\mid|h_1(t)|^2 \ge T, |h_2(t)|^2 \ge T\right\}\\
&\cS_2 = \left\{t \in \{1,n\}\mid|h_1(t)|^2 \ge T, |h_2(t)|^2 < T\right\}\\
&\cS_3 = \left\{t \in \{1,n\}\mid|h_1(t)|^2 < T, |h_2(t)|^2 \ge T\right\}\\
&\cS_4= \left\{t \in \{1,n\}\mid|h_1(t)|^2 < T, |h_2(t)|^2 <
T\right\}.\end{aligned}\label{eq:setDefn}\end{equation}

The resulting channel is a probabilistic parallel channel with probabilities of the four channels as $p(\cS_1)
=\Pr(|H_1|^2\ge T, |H_2|^2\ge T)$, $p(\cS_2) = \Pr(|H_1|^2\ge T, |H_2|^2
< T)$, $p(\cS_3) =\Pr(|H_1|^2 < T, |H_2|^2\ge T)$ and $p(\cS_4) = \Pr(|H_1|^2
<T, |H_2|^2 < T)$. Also note that with $X_j = U_j \sim \CN(0,P)$ in the argument of the summation in~\eqref{eq:ProbabilisticCommonMessageLowerBound},\begin{equation}
\{I(U_j;Y_{ij})-I(U_j;Y_{ej})\}^+ = \begin{cases} 0, & \mrm{if~} |H_i|^2 \le T \text{ in } \cS_j\\
E\left[\log(1 + |H_i|^2 P) - \log(1 + |H_e|^2 P) \mid |H_i|^2 \ge
T\right], & \mrm{if~} |H_i|^2
> T \text{ in } \cS_j.\end{cases}
\end{equation}
Substituting these expressions in the achievable rate expression for the
probabilistic parallel channel~\eqref{eq:ProbabilisticCommonMessageLowerBound}
we get

%
%
%
%The achievable rate is given by:
\begin{align}
R^\mrm{common}(P) &= \max_{T > 0} \min_{1\le i \le 2}\Pr(|H_i|^2 \ge
T) E\left[\log(1 + |H_i|^2 P) - \log(1 + |H_e|^2
P) \Biggm| |H_i|^2 \ge T\right] \nonumber\\
&= \max_{T > 0} \min_{1\le i \le 2} \int_{T}^\infty (\log(1 + xP)
- E[\log(1 + |H_e|^2 P])p_i(x) dx \nonumber\\
&\ge \min_{1\le i \le 2} \int_{T^*}^\infty (\log(1 + xP) - E[\log(1
+ |H_e|^2 P]) p_i(x) dx \label{eq:FadingCommonMsgAchiev}\\
&= \min_{1\le i \le 2} \int_{0}^\infty \{\log(1 + xP) - E[\log(1
+ |H_e|^2 P)]\}^+ p_i(x) dx \nonumber\\
 &=\min_{1\le i \le 2}E\left[\{\log(1+
|H_i|^2 P) - E[\log(1 + |H_e|^2 P)]\}^+\right],\label{eq:finalTerm}
\end{align}
where $T^*$ in~\eqref{eq:FadingCommonMsgAchiev} is the solution to
$\log(1 + xP) - E[\log(1 + |H_e|^2 P]=0$. (The optimality of $T^*$
follows from the fact that $p_i(x)\ge 0$ and hence the integral is
maximized by keeping all terms which are positive and discarding the
negative terms, however this is not necessary to note as this is an achievable scheme.) Note that~\eqref{eq:finalTerm} coincides with the
achievable rate in Theorem~\ref{thm:Wireless_CommonMsg} for the case
of $K=2$ users. As remarked earlier, this scheme straightforwardly
generalizes to more than two receivers. With $K$ receivers we will
have a total of $2^K$ states, where each state specifies the subset
of users that are above the threshold $T^*$.

\section{Proof of the Upper Bound in Lemma~\ref{lem:Wireless_SumCapacityLowerBound}}

\label{app:GenieAidedFadingChannel}

Consider the channel with one receiver and one eavesdropper.
\begin{equation}
\begin{aligned}
Y(t) = H_\mrm{max}(t) X(t) + Z(t)\\
Y_e(t) = H_e(t)X(t) + Z_e(t). \label{eq:genieAidedChannel}
\end{aligned}
\end{equation}Along the lines of Lemma~\ref{lem:GenieAidedChannel} in
Section~\ref{susec:sumCapParChanUB} one deduces that the
sum-secrecy-capacity of the channel~\eqref{eq:FastFading} is upper
bounded by the secrecy capacity of the
genie-aided-channel~\eqref{eq:genieAidedChannel}. It remains to show
that an upper bound on the secrecy capacity of this channel is
\begin{equation}
R^{+}(P) = \max_{P(H_\mrm{max}): E[P(H_\mrm{max})]\le P}
E\left[\{{\log(1+|H_\mrm{max}|^2 P(H_\mrm{max}))}-{\log(1+ |H_e|^2
P(H_\mrm{max}))}\}^+\right]. \label{eq:GenieAidedFadingChannel}
\end{equation}

In what follows we will denote the eavesdropper's channel output by
$\hat{Y}_e(t) = (Y_e(t), H_e(t))$ and optimistically assume that the
sequence $H_\mrm{max}^n$ is known to the sender and the receiver
non-causally. The joint distribution of the noise variables
$(Z(t),Z_e(t))$ is selected to be such that if $|H_e(t)| \le
|H_\mrm{max}(t)|$ we have $X(t) \rightarrow Y(t) \rightarrow
Y_e(t)$, otherwise we have $X(t) \rightarrow Y_e(t) \rightarrow
Y(t)$.

Suppose for this channel and the sequence $H_\mrm{max}^n$, there is
a sequence of $(n,2^{nR})$ codes that achieve perfect secrecy in
 Definition~\ref{def:Wireless_code2}. Following the derivation of the upper bound Theorem~\ref{thm:SumCap}, we have
 \begin{align}
nR &\le I(W;Y^n|H_\mrm{max}^n) - I(W;\hat{Y}_e^n| H_\mrm{max}^n) + n\eps \nonumber\\
&\le I(W;Y^n,\hat{Y_e}^n|H_\mrm{max}^n) - I(W;\hat{Y}_e^n| H_\mrm{max}^n) + n\eps \nonumber\\
&= I(W;Y^n| H_\mrm{max}^n, \hat{Y}_e^n)+ n\eps \nonumber\\
&\le I(X^n;Y^n| H_\mrm{max}^n, \hat{Y}_e^n)+ n\eps \label{eq:FadingSumCapConvMC}\\
&\le \sum_{t=1}^n I(X(t);Y(t)| H_\mrm{max}(t), \hat{Y}_e(t))+ n\eps
\label{eq:FadingSumCapConvMem}
\end{align}
where~\eqref{eq:FadingSumCapConvMC} follows from the fact that $W
\rightarrow (X^n, \hat{Y_e^n}) \rightarrow Y^n$ follows a Markov
chain and~\eqref{eq:FadingSumCapConvMem} from the fact that the channel
is memoryless.

Now let $\cH_n$ be the set of all fades that have been realized,
i.e.,
\begin{equation}
\cH_n = \left\{\g \mid \exists t \in [1,n], H_\mrm{max}(t) =
\g\right\},
\end{equation} let $N_\g$ denote the number of times in the interval $[0,n]$
that the channel has fade $\g$, and let $S_\g$ denote the time indices
corresponding to a fade $\g$, i.e.,
$$S_\g = \left\{t\mid 1\le t \le n, |H_\mrm{max}(t)|^2 =\g\right\}\quad \g \in \cH_n.$$
Letting the average transmitted power at time $t \in \cS_\g$ be denoted
as $P_\g^n(t)$ we have
\begin{equation}
P^n(t) \defeq
 E[|X(t)|^2] \quad t \in \cS_\g
\end{equation}
where the expectation is over the set of transmitted messages and
any stochastic mapping used by the encoder.  Let $\bar{P}_\g^n$
denote the average power transmitted with fade level $\g$
\begin{equation}
\bar{P}_\g^n = 
\begin{cases}
\frac{1}{N_\g}\sum_{t\in\cS_\g} P^n(t),& \g \in
\cH_n\\
0 & \mrm{otherwise}
\end{cases}
\end{equation}

We will need the following Lemma, which follows from the capacity
for the Gaussian wiretap channel~\cite{leung-Yan-CheongHellman:99}.
\begin{lemma}
\label{lem:UpperBoundingHelper} Let $(X,Y,\hat{Y_e})$ be
random variables such that $Y=\sqrt{\g} X + Z_r$ and $Y_e=\sqrt{\mu}
X + Z_r$. Suppose that $Z_r \sim \CN(0,1)$ and $Z_e \sim \CN(0,1)$
and that the joint distribution of $(Z_r, Z_e)$ satisfies $X\rightarrow Y
\rightarrow Y_e$ if $|\mu|\le |\g|$ and $X\rightarrow Y_e
\rightarrow Y$ otherwise. Then we have
\begin{equation}\max_{p(X), E[|X|^2]\le \bar{P}}I(X;Y| Y_e) =
\log(1 + \g \bar{P}) - \log(1 +\min(\g, \mu)\bar{P} )\;.
\end{equation}
\end{lemma}
Now we have
\begin{align}
&\quad\sum_{t=1}^n I\left(X(t);Y(t)\Biggm| H_\mrm{max}(t), \hat{Y}_e(t)\right)\nonumber\\
&=\sum_{\g_0 \in \cH_n} \sum_{t\in \cS_{\g_0}} I\left(X(t);Y(t)\mid
\Biggm|H_\mrm{max}(t)|^2 = \g_0,\hat{Y}_e(t)\right)\label{eq:rewriteSum}\\
&=\sum_{\g_0 \in \cH_n} \sum_{t\in
\cS_{\g_0}}\left\{\int_{\mu=0}^\infty I\left(X(t);Y(t)\Biggm|
|H_\mrm{max}(t)|^2 = \g_0,{Y}_e(t),|H_e(t)|^2=\mu\right) p_\mu d\mu
\right\}\label{eq:indepClaim_pmu}\\
&\le \sum_{\g_0 \in \cH_n} \sum_{t\in \cS_{\g_0}}
\left\{\int_{\mu=0}^\infty
(\log(1 + \g_0 P^n(t)) - \log(1 + \min(\g_0,\mu )P^n(t))) p_\mu d\mu \right\}\label{eq:Ineq1}\\
&\le \sum_{\g_0 \in \cH_n} \sum_{t\in \cS_{\g_0}}
(\log(1 + \g_0 P^n(t)) - E_\mu[\log(1 + \min(\g_0,\mu )P^n(t)))]\nonumber\\
&\le \sum_{\g_0 \in \cH_n} N_{\g_0} (\log(1 + \g_0 \bar{P}_{\g_0}^n)
- E_\mu[\log(1 + \min(\g_0,\mu)\bar{P}_{\g_0}^n)]) \label{eq:Ineq2},
\end{align}
where~\eqref{eq:rewriteSum} follows by re-writing the summation over
the set $\cH_n$,~\eqref{eq:indepClaim_pmu} follows from the independence of $H_e(t)$ and $H_\mrm{max}(t)$,~\eqref{eq:Ineq1} follows from
Lemma~\ref{lem:UpperBoundingHelper}, and~\eqref{eq:Ineq2} is
justified from the fact that for any  $\g_0> 0$, the expression
$\Psi_{\g_0}(y) =\log(1 + \g_0 y) - E[\log(1 + \min(\g_0,\mu)y)]$ is
a concave function in $y$ for $y\ge 0$, i.e., $$\sum_{t\in
\cS_{\g_0}}\Psi(P_{\g_0}^n(t)) \le N_{\g_0}
\Psi\left(\frac{1}{N_{\g_0}}\sum_{t\in
\cS_{\g_0}}P_{\g_0}^n(t)\right)=N_{\g_0} \Psi(\bar{P}_{\g_0}^n).$$
Combining ~\eqref{eq:FadingSumCapConvMem} and~\eqref{eq:Ineq2} we get
\begin{equation}
R \le \eps + \sum_{\g_0 \in \cH_n}\log(1 + \g_0\bar{P}_{\g_0}^n) -
E_\mu[\log(1 +\min(\g_0, \mu)\bar{P}_{\g_0}^n )] \frac{N_{\g_0}}{n}.
\label{eq:SumCapUpperBoundN}
\end{equation}
Now note that, for each $n$, we have $\int_{\g=0}^\infty
\bar{P}_{\g_0}^n d\g \le P$. Thus for each $n$, the set of points $\bar{P}^n_{\g_0}$, indexed by $\g_0$, lie
in a compact space. For this sequence of points, there exists a
convergent subsequence $\bar{P}^{n_i}_{\g_0}$ that converges to some power
allocation $\bar{P}_{\g_0}$ as $n_i \rightarrow \infty$. Taking limit along
the converging subsequence of the upper bound on the
rate~\eqref{eq:SumCapUpperBoundN}
\begin{align*}
R &\le  \int_{0}^\infty \log(1 + \g\bar{P}_\g) - E[\log(1
+\min(\g, \mu)\bar{P}_\g )] p_\g d\g\\
&= E[\log(1 + \g\bar{P}_\g) - \log(1 +\min(\g, \mu)\bar{P}_\g )],
\end{align*}
and this completes the proof of the upper bound.

\section{Helper Lemma in the proof of Theorem~\ref{thm:WirelessAsymptoticSumCapacity}}
\label{app:WirelessAsymptoticSumCapacity}

\begin{lemma}
Let $H_1, H_2,\ldots, H_K, H_e$ be i.i.d.\ unit mean exponentials.  For $K\ge
2$, we have
$$E\left[\log\frac{|H_e|^2}{|H_\mrm{max}|^2}\Biggm | |H_e|^2\ge
|H_\mrm{max}|^2\right] \le 2 \log 2$$
\label{lem:OrderedStatisticsBound}
\end{lemma}

First note the following.
\begin{fact}[~\cite{David}]
Let $V_1,V_2,\ldots, V_K,V_{K+1}$ be i.i.d. exponential random
variables with mean $\lambda$ and ${V_\mrm{max}(K+1)}$ denotes the
largest of these exponential and ${V_\mrm{max}(K)}$ the second
largest. The joint distribution of
$({V_\mrm{max}(K)},{V_\mrm{max}(K+1)})$ satisfies
\begin{equation}
{V_\mrm{max}(K+1)} = {V_\mrm{max}(K)} + Y, \end{equation} where $Y$
is an exponential random variable with mean $\lambda$ and is
independent of ${V_\mrm{max}(K)}$ \label{fact:ExpRV}
\end{fact}

\begin{proof}
We have
\begin{align}
E\left[\log\frac{|H_e|^2}{|H_\mrm{max}|^2}\Biggm | |H_e|^2\ge
|H_\mrm{max}|^2\right] &= E\left[\log\frac{|H_\mrm{max}|^2 +
Y}{|H_\mrm{max}|^2}\right] \\
&\le E\left[\frac{
Y}{|H_\mrm{max}|^2}\right]\label{eq:logUB}\\
&=E[Y]E\left[\frac{ 1}{|H_\mrm{max}|^2}\right] \label{eq:indep}\\
&=E\left[\frac{ 1}{|H_\mrm{max}|^2}\right] \label{eq:indep2}
\end{align}
where~\eqref{eq:logUB} follows from the identity
$\log(1+x)\le x$ for $x>0$,~\eqref{eq:indep} follows from the
independence of $Y$ and $H_\mrm{max}$, and~\eqref{eq:indep2} from
the fact that $E[Y]=1$. Since $|H_\mrm{max}|^2 \ge
\max(|H_1|^2,|H_2|^2)$ we obtain
$$E\left[\frac{1}{|H_\mrm{max}|^2}\right] \le E\left[\frac{1}{\max(|H_1|^2,|H_2|^2)}\right] \le 2\log
2.$$

\end{proof}

\bibliographystyle{IEEEtranS}
\bibliography{references}
\end{document}